\documentclass[10pt,twoside]{article}
\usepackage{capt-of}
\usepackage{amsfonts}
\usepackage{amssymb}
\usepackage{amsmath}
\usepackage[dvips]{graphicx}
\usepackage[latin1]{inputenc}
\usepackage{slashed}
\usepackage{pifont}
\usepackage{float}
\usepackage{graphicx}
\usepackage{multirow}
\usepackage{authblk}

\providecommand{\NEG}{\slashed}

\begin{document}

\title{Constraints on neutrino masses coming from magnetic dipole moments in
a two Higgs doublet model type I and II}
\author{Carlos G. Tarazona\thanks{%
caragomezt@unal.edu.co}$^{1,2}$, Rodolfo A. Diaz\thanks{%
radiazs@unal.edu.co}$^{1}$, John Morales\thanks{%
jmoralesa@unal.edu.co}$^{1}$. \\
%EndAName
$^{1}$Departamento de Física. Universidad Nacional de Colombia. Bogotá,
Colombia.\\
$^{2}$Departamento de Ciencias Básicas. Universidad Manuela Beltrán. Bogotá,
Colombia.}
\date{}
\maketitle

\begin{abstract}
In the framework of a two Higgs doublet model type I and type II, we
calculate limits on neutrino masses for the different types of neutrinos, by
using the experimental bounds on their magnetic dipole moments. This is
carried out by analyzing diagrams of Cherenkov neutrino decays with a
charged Higgs into the loop, coming \ from the two Higgs doublet model
(2HDM). Such constraints are translated into allowed regions in the free
parameters of the models, for each neutrino flavor.

The analysis was performed by sweeping the charged Higgs mass between $%
(100-900)GeV$ and taking into account the experimental constraints for $\tan
\beta $ in the 2HDM type I and II, obtaining contributions close to the
experimental thresholds for muon and tau neutrinos, while for electron
neutrino the relevant contribution comes from standard model and keeps out
of the reach of forthcoming experiments.

{\small \textbf{PACS:} 41.20.Cv, 02.10.Yn, 01.40.Fk, 01.40.gb, 02.30.Tb }
\end{abstract}

\section{Introduction}

The phenomenon of neutrino oscillations, has been supported by the discovery
of flavor conversions of neutrinos from different sources, like the
atmospheric neutrinos made by Super-Kamiokande in 1998\cite{Fukuda}, or more
recently by T2K Collaboration\cite{Abe}. These oscillations occur among at
least three types of flavors of neutrinos: electron, muon and tau neutrinos.%
\begin{equation}
\left\vert \nu _{\alpha }\right\rangle =\sum\limits_{k}U_{\alpha k}^{\ast
}\left\vert \nu _{k}\right\rangle \text{ \ \ \ \ \ \ \ }\left( \alpha
=e,~\mu ,~\tau \right)
\end{equation}%
all neutrinos produced and observed so far, have left-handed helicities,
while all antineutrinos have right-handed helicities. Neutrino oscillations
provided the first glimpse of physics beyond the standard model of
particles. Now, since neutrino oscillations are sensitive only to the
difference in the squares of their masses, such a phenomenon requires that
at least two neutrino species have nonzero mass. The transition probability
between different flavors can be approximated by%
\begin{equation}
P_{\nu _{\alpha }\rightarrow \nu _{\beta }}\left( t\right)
=\sum\limits_{k,j}U_{\alpha k}^{\ast }U_{\beta k}U_{\alpha j}U_{\beta
j}^{\ast }\exp \left( -i\frac{\Delta m_{kj}^{2}L}{2E}\right)
\end{equation}%
where $P_{\nu _{\alpha }\rightarrow \nu _{\beta }}\left( t\right) $ is the
probability that after travelling a distance $L$, a neutrino with flavor $%
\nu _{\alpha }$ converts into a neutrino with flavor $\nu _{\beta }$. As for
the neutrino masses, we have only upper bounds hitherto\cite{PDG}%
\begin{equation}
m_{\nu _{e}}\leq 2.05eV\ ,\ \ m_{\nu _{\mu }}\leq 0.19MeV\ ,\ \ m_{\nu
_{\tau }}\leq 18.2MeV
\end{equation}

All elementary fermions in the Standard Model are Dirac fermions.
Nevertheless, the nature of the neutrino is not yet definitely settled and
depending on the model the neutrino can be either a Majorana or Dirac
fermion. On the other hand, despite the neutrinos do not carry electric
charge, they can participate in electromagnetic interactions by coupling
with photons via loop diagrams, and like other particles the electromagnetic
properties can be described by electromagnetic form factors (EFF's). For
example, by means of its multipole moments, neutrinos can be sensitive to
intense electromagnetic fields, and such intense fields can exist in nature.
It has been suggested that there could be sources of magnetic fields of
order $\left( 10^{13}-10^{18}\right) G$, as it could be the case during a
supernova explosion or in the vicinity of special groups of neutron stars
known as magnetars\cite{Mereghetti}.

On the other hand, present limits on the scalar sector in the standard
model, still permits the possibility of an extended Higgs sector. We shall
study one of the simplest extension of the scalar sector of the standard
model, the so-called Two Higgs Doublet Model (2HDM) in which we add a second
Higgs doublet with the same quantum numbers of the first. There are many
motivations for this model, one of this is the fact that the SM is unable to
generate a baryon asymmetry of the universe of sufficient size, or to
explain the mass hierarchy in the third generation of quarks. Two Higgs
Doublet models are possible scenarios to solve these problems, due to the
flexibility of their scalar mass spectrum and the existence of additional
sources of $CP$ violation. In addition in the Minimal Supersymmetric
Standard Model (MSSM), a second doublet should be added in order to cancel
anomalies\cite{Sher}.

The coupling of neutrinos with photons occur via loop diagrams. In the
Standard Model (SM), the loop corrections have the form of vertex diagrams
and vacuum polarization diagrams. When a second doublet of scalars is
included in the spectrum, further corrections appear by replacing the vector
bosons $W^{\pm }$ by charged Higgs bosons $H^{\pm }$. Our goal is to
characterize the corrections to the EFF's coming from the new physics, and
particularly on the region of parameters in which such factors become near
the threshold of detection. In the region of parameters in which the
threshold of detection is reached, we obtain bounds on neutrino masses.

The structure of this paper is as follows. In section \ref{Dirac masses} we
discuss briefly the implementation of neutrino Dirac masses in SM. In
section \ref{form factors} we discuss the general form of the EFF's for
neutrinos. In section \ref{2HDM} we describe briefly the two Higgs doublet
Model (2HDM), particularly the models of type I and of type II as well as
the implementation of neutrinos masses into those models. In section \ref%
{loops}, we characterize the loop diagrams coming from the 2HDM that
contributes to the EFF's of the neutrino. In section \ref{results} we find
upper bounds for the neutrino masses in the framework of the 2HDM type I and
II, by using the allowed values of the free parameters of the model, as well
as the experimental limits for the magnetic dipole moments of such
neutrinos. Finally, section \ref{sec:conclusions} yields our conclusions.

\section{Neutrino Dirac mass term in SM\label{Dirac masses}}

There are several ways to incorporate neutrino masses within the SM or its
extensions, in order to explain the observed neutrino oscillations. We shall
use a simple form which consists of adding right-handed singlets of
neutrinos fields $\left( \nu _{\alpha ^{\prime }R}\right) $ corresponding to
each charged lepton. This insertion implies new gauge invariant interactions
in the Yukawa sector%
\begin{equation}
-\mathcal{L}_{Yukawa}=\sum\limits_{\alpha =e,\mu ,\tau }\sum\limits_{\alpha
^{\prime }=1}^{3}f_{\alpha ,\alpha ^{\prime }}\overline{\psi }_{\alpha L}%
\widetilde{\Phi }\nu _{\alpha ^{\prime }R}+h.c  \label{Yukawa neutrino}
\end{equation}%
where $f_{\alpha ,\alpha ^{\prime }}$ is a matrix with new coupling
constants, $\psi _{\alpha L}$ is the left-handed lepton doublet and $\Phi $
is the SM\ Higgs doublet, with\ $\widetilde{\Phi }\equiv i\sigma _{2}\Phi
^{\ast }$ 
\begin{equation}
\left\langle \Phi \right\rangle =\frac{1}{\sqrt{2}}\left( 
\begin{array}{c}
0 \\ 
v%
\end{array}%
\right)
\end{equation}%
a nonzero vacuum expectation value (VEV)\ of the Higgs doublet induces the
spontaneous symmetry breaking from $SU\left( 2\right) _{L}\times U\left(
1\right) _{Y}$ to $U\left( 1\right) _{Q}$. In turn, the VEV also provides
the neutrino Dirac mass term%
\begin{equation}
-\mathcal{L}_{D_{mass}}=\frac{v}{\sqrt{2}}\sum\limits_{\alpha =e,\mu ,\tau
}\sum\limits_{\alpha ^{\prime }=1}^{3}f_{\alpha ,\alpha ^{\prime }}\overline{%
\nu }_{\alpha L}\nu _{\alpha ^{\prime }R}+h.c
\end{equation}%
In general the matrix $f_{\alpha ,\alpha ^{\prime }}$ is a complex $3\times
3 $ matrix, the massive neutrino fields are obtained through the
diagonalization of $\mathcal{L}_{D_{mass}}$ , this can be done by
diagonalizing $f_{\alpha ,\alpha ^{\prime }}$ with a biunitary transformation%
\begin{equation}
\frac{v}{\sqrt{2}}\left( U^{\dagger }~f~V\right) _{k,j}=m_{k}\delta _{kj}
\end{equation}%
where $m_{k}$ is a diagonal matrix with real and positive values.

The flavor eigenstates $\nu _{e},~\nu _{\mu }$ and $~\nu _{\tau }$ are
superpositions of the mass eigenstates%
\begin{equation*}
\nu _{\alpha L}=\sum\limits_{k=1}^{3}U_{\alpha ,k}\nu _{kL}\ \ ;\ \ \nu
_{\alpha ^{\prime }R}=\sum\limits_{j=1}^{3}V_{\alpha ^{\prime },j}\nu _{jR}
\end{equation*}%
or equivalently, the mass eigenstates $\nu _{k}$ are mixtures of flavor
eigenstates. The mixing matrix $U$ is called the PMNS matrix due to
Pontecorvo\cite{Pontecorvo}, Maki, Nakagawa and Sakata\cite{Maki}.

The result of the diagonalization gives Dirac mass terms of the form%
\begin{equation}
-\mathcal{L}_{D_{mass}}=\sum\limits_{k=1}^{3}m_{k}\overline{\nu }_{k}\nu _{k}
\end{equation}%
with the Dirac fields of massive neutrinos given by 
\begin{equation}
\nu _{k}=\nu _{kL}+\nu _{kR}
\end{equation}

\section{The electromagnetic form factors\ (EFF's)\label{form factors}}

\begin{figure}[!htbp]
\centering
\includegraphics[scale=1.1]{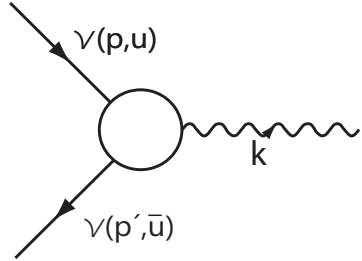} \vspace{0.3cm}
\caption{Effective coupling of two neutrinos with a photon}
\label{vertex}
\end{figure}

\begin{equation}
\left\langle u\left( p,\lambda \right) |J_{\mu }^{EM}\left( x\right)
|u\left( p^{\prime },\lambda ^{\prime }\right) \right\rangle =\overline{u}%
\left( p,\lambda \right) \Lambda _{\mu }\left( l,q\right) u\left( p^{\prime
},\lambda ^{\prime }\right)
\end{equation}%
To find all the EFF's, we use the general expression for the current\cite%
{Nova},\cite{broggini}, where $q_{\mu }=p_{\mu }^{\prime }-p_{\mu }$, $%
l_{\mu }=p_{\mu }^{\prime }+p_{\mu }$ are the four-momenta shown in Fig. \ref%
{vertex}, and\ $u\left( p,\lambda \right) $, $u\left( p^{\prime },\lambda
^{\prime }\right) $ are the initial and final fermion states respectively.
Further, $\Lambda _{\mu }$ are matrices of couplings acting on the spinors.
The matrices $\Lambda _{\mu }$ have some interesting properties

\begin{itemize}
\item The first condition is that the arrangement $\Lambda _{\mu }$ must be
a 4-vector, i.e. must be Lorentz covariant.

\item The second condition is hermiticity of the associated current, i.e., $%
J_{\mu }^{^{\dag }EM}=J_{\mu }^{EM}$ which implies%
\begin{equation}
\Lambda _{\mu }\left( l,q\right) =\gamma ^{0}\Lambda _{\mu }^{\dag }\left(
l,-q\right) \gamma ^{0}
\end{equation}

\item The current conservation or gauge invariance $\partial ^{\mu }J_{\mu
}^{EM}=0$ can be recast into%
\begin{equation}
q^{\mu }\overline{u}\left( p^{\prime },\lambda ^{\prime }\right) \Lambda
_{\mu }\left( l,q\right) u\left( p,\lambda \right) =0
\end{equation}
\end{itemize}

The most general expression for $\Lambda _{\mu }\left( l,q\right) $ reads%
\cite{Giunti}%
\begin{eqnarray}
\Lambda _{\mu }\left( q\right) &=&F_{Q}\left( q^{2}\right) \gamma _{\mu }+ 
\left[ F_{M}\left( q^{2}\right) i+F_{E}\left( q^{2}\right) \gamma _{5}\right]
\sigma _{\mu \nu }q^{\nu } + F_{A}\left( q^{2}\right) \left( q^{2}\gamma
_{\mu }-q_{\mu }\NEG{q}\right) \gamma _{5}
\end{eqnarray}%
where $F_{Q},~F_{M},~F_{E}$ and $F_{A}$ represent the electric charge,
dipole magnetic moment, dipole electric moment and anapole moment
respectively.

The EFF's show us how the particles are coupled with the photon at the tree
level or in loop corrections. At the tree level we got the electric charge
and one part of the contribution coming from the magnetic dipole moment.
Now, if we consider the interaction with an external field $A_{ext}^{\mu }$
in the form 
\begin{equation}
\mathcal{L}_{ext}=-eA_{ext}^{\mu }J_{\mu }^{EM}
\end{equation}%
the so-called anomalous magnetic moment arises. Even uncharged particles may
have magnetic dipole moment. However, for uncharged particles all dipole
moments only appear in loop corrections. Just like the anomalous magnetic
moment, the dipole electric moment and the anapole moment can be non-zero
even for an uncharged particle\cite{lepton}. We summarize some
electromagnetic properties of charged leptons in table \ref{tab:EMFC} 
\begin{table}[!htbp]
\begin{center}
\begin{equation*}
\begin{tabular}{|l|l|l|l|}
\hline
$l$ & Mass$\left( \frac{MeV}{c^{2}}\right) $ & MDM & EDM$\left( \frac{e}{%
2m_{l}}\right) $ \\ \hline
$e$ & $0.51$ & $1.159\times 10^{-3}\mu _{B}$ & $<1\times 10^{-16}$ \\ \hline
$\mu $ & $105.658$ & $1.159\times 10^{-3}\frac{e}{2m_{\mu }}$ & $<2\times
10^{-6}$ \\ \hline
$\tau $ & $1.776\times 10^{3}$ & $1.159\times 10^{-3}\frac{e}{2m_{\tau }}$ & 
$<2\times 10^{-2}$ \\ \hline
\end{tabular}%
\end{equation*}%
\end{center}
\caption{Electromagnetic properties of charged leptons, MDM represents the
magnetic dipole moment and EDM electric dipole moment}
\label{tab:EMFC}
\end{table}

Like other particles, neutrinos can be described by EFF's with vertex
functions. For neutrinos the magnetic and electric dipole moments are
expected to be very small since they are likely proportional to the neutrino
masses. For the anomalous magnetic moment the leading contribution is \cite%
{lepton}%
\begin{equation}
a_{\nu _{i}}=-\frac{3G_{F}m_{\nu _{i}}}{4\sqrt{2}\pi ^{2}}m_{e}
\end{equation}%
Consequently, the neutrino magnetic moment is \cite{Fujikawa}%
\begin{eqnarray}
\vec{\mu}_{\nu _{i}} &=&-\frac{e}{m_{e}}a_{\nu _{i}}\vec{s}_{\nu _{i}} \\
&\Rightarrow &\mu _{\nu }=-\frac{3G_{F}em_{\nu _{i}}}{4\sqrt{2}\pi ^{2}}%
\simeq 3.2\times 10^{-19}\left( \frac{m_{\nu }}{1eV}\right) \mu _{B}  \notag
\end{eqnarray}%
where $\mu _{B}$ is the Bohr's magneton. If neutrino couples to photons via
such moments, the neutrino electromagnetic properties can be used to
distinguish Majorana and Dirac neutrinos. For Dirac neutrinos the most
relevant moment is $F_{M}$, because the other terms vanish in a $CP-$%
conserving scenario with an hermitian $J_{\mu }^{EM}$, and are highly
supressed owing to the soft violation of $CP$. On the other hand, for
Majorana neutrinos only $F_{A}$ is possible, because the other terms vanish
owing to the self-conjugate nature of Majorana neutrinos. Table \ref%
{tab:EMFN} summarizes the electromagnetic properties of massive neutrinos%
\cite{lepton} 
\begin{table}[!htbp]
\begin{center}
\begin{equation*}
\begin{tabular}{|l|l|l|}
\hline
$l$ & Mass$\left( \frac{MeV}{c^{2}}\right) $ & Magnetic dipole moment \\ 
\hline
$\nu _{e}$ & $<2.2\times 10^{-6}$ & $<10.8\times 10^{-10}\mu _{B}$ \\ \hline
$\nu _{\mu }$ & $<0.17$ & $<7.4\times 10^{-10}\mu _{B}$ \\ \hline
$\nu _{\tau }$ & $<16$ & $<5.4\times 10^{-7}\mu _{B}$ \\ \hline
\end{tabular}%
\end{equation*}%
\end{center}
\caption{Electromagnetic properties of massive neutrinos.}
\label{tab:EMFN}
\end{table}

\section{The two Higgs doublet model with massive neutrinos\label{2HDM}}

The 2HDM contains five Higgs bosons in its spectrum\cite{2HDM}. The symmetry
breaking is implemented by introducing a new scalar doublet with the same
quantum numbers of the first one\cite{Diaz}. In a CP-conserving scenario,
the Higgs sector consists of: Two Higgs CP-even scalars $\left(
H^{0},h^{0}\right) ,$ one CP-odd scalar $\left( A^{0}\right) $ and two
charged Higgs bosons $\left( H^{\pm }\right) $. A key parameter of the model
is the ratio between the vacuum expectation values%
\begin{equation}
\tan \beta =\frac{v_{2}}{v_{1}}
\end{equation}%
where $v_{1}$ and $v_{2}$ are the vacuum expectation values of the Higgs
doublets\cite{Carena}, with values of $0\leq \beta \leq \frac{\pi }{2}$.

The most general gauge invariant Lagrangian that couples the Higgs fields to
leptons (with massless neutrinos) reads 
\begin{equation}
-\mathcal{L}_{Y}=\eta _{i,j}^{E,0}\overline{l}_{iL}^{0}\Phi
_{1}E_{jR}^{0}+\xi _{i,j}^{E,0}\overline{l}_{iL}^{0}\Phi _{2}E_{jR}^{0}+h.c.
\notag
\end{equation}%
where $\Phi _{1,2}\;$represents the Higgs doublets, and$\ \widetilde{\Phi }%
_{1,2}\equiv i\sigma _{2}\Phi _{1,2}$, The superscript \textquotedblleft $0$%
\textquotedblright\ indicates that the fields are not mass eigenstates yet,$%
\;\eta _{i,j}\;$and $\xi _{i,j}\;$are non diagonal $3\times 3$ matrices with 
$\left( i,j\right) \;$denoting family indices. $E_{jR}^{0}\;$denotes the
three charged leptons and$\ \overline{l}_{iL}^{0}\;$denotes the lepton weak
isospin left-handed doublets.

It is customary to implement a discrete symmetry in the 2HDM in order to
suppress some processes such as the Flavor Changing neutral currents (FCNC).
In particular by demanding the discrete symmetry 
\begin{eqnarray}
\Phi _{1} &\rightarrow &\Phi _{1}\ \ ;\ \ \ \Phi _{2}\rightarrow -\Phi _{2} 
\notag \\
D_{jR} &\rightarrow &\mp D_{jR}\ \ ;\ \ U_{jR}\rightarrow -U_{jR}
\label{discrete sym}
\end{eqnarray}%
such kind of processes are eliminated at the tree-level. Here $D_{jR}$ and$\
U_{jR}$ denote right-handed singlets of the down and up types of fermions.

\subsection{The 2HDM type I}

By taking $D_{jR}\rightarrow -D_{jR}$ we arrive to the so-called 2HDM of
type I. In this scenario, only $\Phi _{2}\;$couples in the Yukawa sector and
gives masses to all fermions. The Lepton Yukawa Lagrangian becomes%
\begin{equation}
-\mathcal{L}_{Y}=\eta _{ij}^{E,0}\overline{l}_{iL}^{0}\widetilde{\Phi }%
_{2}\nu _{jR}^{0}+\xi _{ij}^{E,0}\overline{l}_{iL}^{0}\Phi
_{2}E_{jR}^{0}+h.c.
\end{equation}%
and the term of charged current of the Lagrangian with leptons yields%
\begin{equation}
-\mathcal{L}_{Y}=\frac{g\cot \beta }{\sqrt{2}M_{W}}\overline{l}\left(
M_{l}^{diag}P_{R}-M_{\nu }^{diag}P_{L}\right) \nu H^{+}+h.c.
\label{type I expand}
\end{equation}

\subsection{The 2HDM type II}

If we use $D_{jR}\rightarrow D_{jR}\;$\ we obtain the so-called 2HDM of type
II. In this model $\Phi _{1}$ couples and gives masses to the down sector,
while $\Phi _{2}$ couples and gives masses to the up sector. In consequence,
the Lepton Yukawa Lagrangian (with massless neutrinos) becomes%
\begin{equation}
-\mathcal{L}_{Y}=\eta _{ij}^{D,0}\overline{l}_{iL}^{0}\widetilde{\Phi }%
_{1}E_{jR}^{0}+\xi _{ij}^{\nu ,0}\overline{l}_{iL}^{0}\Phi _{2}\nu
_{jR}^{0}+h.c.  \label{type II}
\end{equation}%
and the term of charged current of the lagrangian with leptons gives%
\begin{equation*}
\mathcal{L}_{Y}=\frac{g}{\sqrt{2}M_{W}}\overline{\nu }\left[ \left( \cot
\beta M_{\nu }^{diag}P_{L}+\tan \beta M_{l}^{diag}P_{R}\right) \right]
lH^{+}+h.c.
\end{equation*}

An interesting aspect is that the limits for the parameter space\ ($%
m_{H^{+}},\tan \beta $), for model type II are very similar to those
obtained by considering the minimal supersymmetric scenario.

\subsection{2HDM type I and II with massive neutrinos}

The term (\ref{Yukawa neutrino}) inserted in the Yukawa sector of the
standard model should also be inserted in the two Higgs doublet model for
each doublet $\Phi _{i}$. Nevertheless, when we implement the discrete
symmetry (\ref{discrete sym}) in a Lagrangian of the form (\ref{Yukawa
neutrino}) we observe that the term involving the doublet $\Phi _{1}$ cannot
appear, and that the extra term is the same in either model type I or model
type II.

\section{Radiative corrections in 2HDM\label{loops}}

\begin{figure}[!htbp]
\centering
\includegraphics[scale=1]{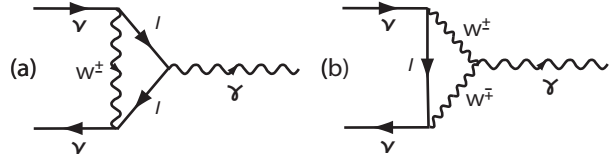} \vspace{0.3cm}
\caption{Loop corrections with leptons and $W^{\pm }$ vector bosons in SM.}
\label{fig3}
\end{figure}

The diagrams that contribute to the neutrino electromagnetic vertex in SM
are displayed in Fig. \ref{fig3} 

\begin{figure}[!htbp]
\centering
\includegraphics[scale=1]{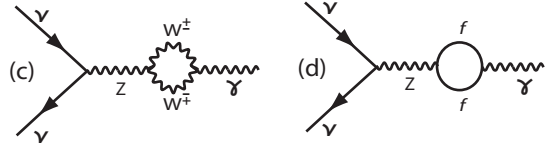} \vspace{0.3cm}
\caption{Vacuum polarization with $W^{\pm }$ vector bosons, and fermions
denoted by $f$ in SM.}
\label{fig4}
\end{figure}

\begin{figure}[!htbp]
\centering
\includegraphics[scale=1]{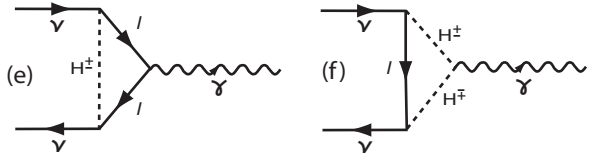} \vspace{0.3cm}
\caption{Loop corrections with leptons and $H^{\pm }$ in the 2HDM. }
\label{fig5}
\end{figure}

\begin{figure}[!htbp]
\centering
\includegraphics[scale=1]{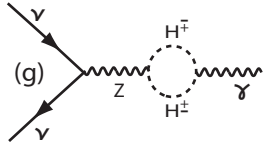} \vspace{0.3cm}
\caption{Vacuum polarization with $H^{\pm }$ in the 2HDM.}
\label{fig6}
\end{figure}

And for the vacuum polarization they are shown in Fig. \ref{fig4}. 
Within the framework of a 2HDM with massive neutrinos, we should add three
new types of diagrams: two vertex corrections illustrated in Fig. \ref{fig5}%
, and one correction to the vacuum polarization displayed in Fig. \ref{fig6}%
. They arise by replacing $W^{\pm }$ by $H^{\pm }$ in the SM diagrams.


Fig. \ref{fig5}(f) shows the vertex correction involving two charged Higgs
bosons and one charged lepton into the loop $\left( 2H^{\pm }1L\right) $.
For this diagram, the general form of the contribution can be written as 
\begin{equation*}
\Lambda _{2H^{\pm }1L}^{\alpha }\left( q,l\right) =-e\int \frac{d^{4}k}{%
\left( 2\pi \right) ^{4}}\left\{ \frac{\left( 2k^{\alpha }+p_{2}^{\alpha
}+p_{1}^{\alpha }\right) \left( aP_{L}+bP_{R}\right) \left( \NEG%
{k}+m_{l}\right) \left( cP_{L}+dP_{R}\right) }{\left[ \left( k+p_{1}\right)
^{2}-m_{H^{\pm }}^{2}\right] \left[ \left( k+p_{2}\right) ^{2}-m_{H^{\pm
}}^{2}\right] \left( k^{2}-m_{l}^{2}\right) }\right\}
\end{equation*}%
where $a,b,c$ and $d$ are constants associated with the Feynman rules of the
2HDM, with $P_{R,L}=\left( 1\pm \gamma ^{5}\right) /2$. The contribution to
the EFF's of this diagram is described in the appendix, and in particular,
the contribution to the magnetic dipole moment (MDM) is given by 
\begin{equation*}
\Lambda _{2H^{\pm }1L}^{\alpha }\left( q,l\right) _{MDM}=\frac{-ie}{16\pi
^{2}}\int\limits_{0}^{1}dx\int\limits_{0}^{x}dy\frac{1}{P^{2}}\left[ \left(
m_{\nu }-\frac{1}{2}m_{l}\right) +\left( m_{l}-3m_{\nu }\right)
x+2x^{2}m_{\nu }\right] i\sigma ^{\alpha \mu }q_{\mu }\left[ \left(
ac+bd\right) +\left( bd-ac\right) \gamma _{5}\right]
\end{equation*}

On the other hand, the diagram in Fig. \ref{fig5}(e) with two leptons and
one charged Higgs into the loop $\left( 2L1H^{\pm }\right) $, gives a
contribution of the form 
\begin{equation*}
\Lambda _{2L1H}^{\alpha }\left( q,l\right) =-e\int \frac{d^{4}k}{\left( 2\pi
\right) ^{4}}\left\{ \frac{\left( aP_{L}+bP_{R}\right) \left( \NEG{k}+\NEG%
{p}_{1}+m_{l}\right) \gamma ^{\alpha }\left( \NEG{k}+\NEG{p}_{2}+m_{l}%
\right) \left( cP_{L}+dP_{R}\right) }{\left[ \left( k+p_{1}\right)
^{2}-m_{l}^{2}\right] \left[ \left( k+p_{2}\right) ^{2}-m_{l}^{2}\right]
\left( k^{2}-m_{H^{\pm }}^{2}\right) }\right\}
\end{equation*}%
from which we obtain the contribution of this diagram to the MDM, that is
given by 
\begin{equation*}
\Lambda _{2L1H^{\pm }}^{\alpha }\left( q,l\right) _{MDM}=\frac{-ie}{16\pi
^{2}}\int\limits_{0}^{1}dx\int\limits_{0}^{x}dy\frac{1}{P^{2}}\left(
2x^{2}m_{\nu }+m_{l}x-xm_{\nu }\right) i\sigma ^{\alpha \mu }q_{\mu }\left[
\left( ac+bd\right) +\left( bd-ac\right) \gamma _{5}\right]
\end{equation*}%
On the other hand the contribution of the vacuum polarization vanishes.
Therefore, the full contribution to the MDM yields%
\begin{equation}
\Lambda _{2HDM}^{\alpha }\left( q,l\right) _{MDM}=\Lambda _{2H^{\pm
}1L}^{\alpha }\left( q,l\right) _{MDM}+\Lambda _{2L1H^{\pm }}^{\alpha
}\left( q,l\right) _{MDM}
\end{equation}%
for the 2HDM type I, the values of $a,b,c$ and $d$ are 
\begin{equation*}
a=c=\frac{2^{\frac{3}{4}}\sqrt{G_{F}}}{\tan \beta }m_{v_{l}}\ \ ;\ \ b=d=%
\frac{2^{\frac{3}{4}}\sqrt{G_{F}}}{\tan \beta }m_{l}
\end{equation*}%
we shall use the numerical value%
\begin{equation*}
G_{F}=\frac{\sqrt{2}}{8}\frac{g^{2}}{M_{W}^{2}}=1.1663787(6)\times
10^{-5}GeV^{-2}
\end{equation*}%
as for the 2HDM type II, the values of $a,b,c$ and $d$ are given by 
\begin{equation*}
a=c=\frac{2^{\frac{3}{4}}\sqrt{G_{F}}}{\tan \beta }m_{v_{l}}\ ;\ \ b,d=2^{%
\frac{3}{4}}\sqrt{G_{F}}m_{l}\tan \beta
\end{equation*}

\section{Results and analysis\label{results}}

Our analysis will be based on constraints on charged Higgs masses and the $%
\tan \beta $ parameter. For either model type I or II the experimental
constraints on the possible values in the $\left( m_{H^{\pm }},\tan \beta
\right) $ parameter space comes from processes such as $B_{u}\rightarrow
\tau \nu _{\tau },~D_{s}\rightarrow \tau \nu _{\tau },~B\rightarrow D\tau
\nu _{\tau },~K\rightarrow \mu \nu _{\mu }$ and $BR\left( B\rightarrow
X_{s}\gamma \right) $\cite{Akeroyda}.

Based on the phenomenological constraints on the 2HDM type I, we take values
of $\tan \beta $ between $\left( 2-90\right) $ and values of the charged
Higgs mass of $m_{H^{\pm }}=\left( 100-300-500-700-900\right) GeV\ $\cite%
{Mahmoudi}. On the other hand, for the 2HDM type II, we have different
allowed intervals of $\tan \beta $ for different values of the charged Higgs
mass: for $m_{H^{\pm }}=300GeV$ the values of $\tan \beta $ lie within the
interval $\left( 4-40\right) $, for $m_{H^{\pm }}=500GeV$ the value of $\tan
\beta $ is between $\left( 2-69\right) $ and for $m_{H^{\pm }}\left(
700-900\right) GeV$ the values of $\tan \beta $ is between $\left(
1-70\right) \ $\cite{Mahmoudi}.

We shall make contourplots of the neutrino mass versus MDM of the neutrino
for different values of the charged Higgs mass sweeping all allowed values
of $\tan \beta $ for each mass. As for the neutrino masses, we shall plot up
to an order of magnitude higher than the upper bound of the SM.

\begin{itemize}
\item Electron neutrino case
\end{itemize}

Taking into account the upper experimental bound in the SM for the electron
neutrino mass$\ m_{\nu _{e}}$, we shall plot within the interval $1\times
10^{-8}MeV\leq m_{\nu _{e}}\leq 1\times 10^{-5}MeV$. If we use the above
interval in a model with two Higgs doublets, for different values of Higgs
mass and $\tan \beta $, we shall obtain exclusion regions by taking as
reference the experimental thresholds for the MDM of the electron neutrino.
In this way, it is possible to obtain upper bounds on the neutrino mass in
this scenario.

\begin{figure}[!htbp]
\centering
\includegraphics[width=0.49\columnwidth]{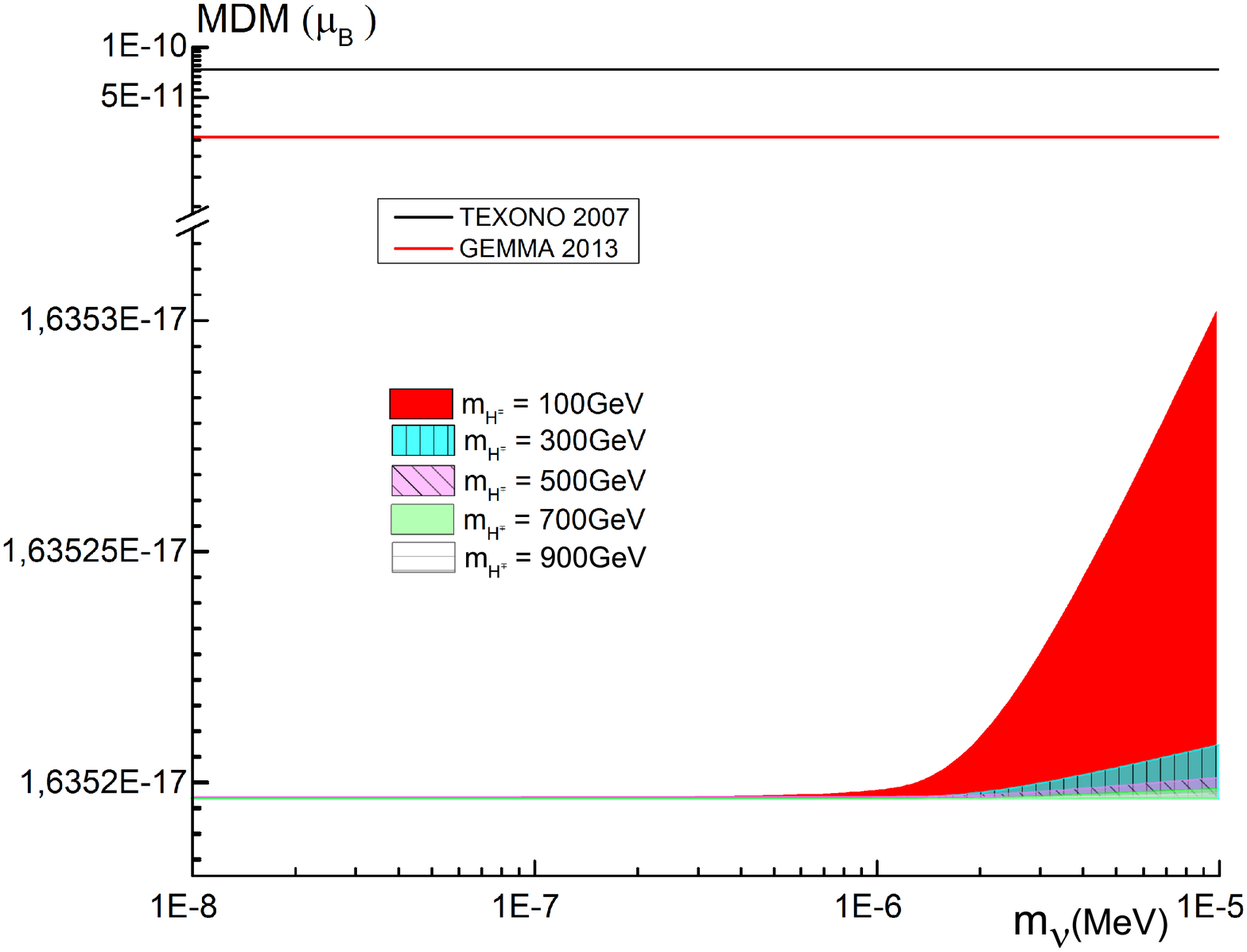} %
\includegraphics[width=0.49\columnwidth]{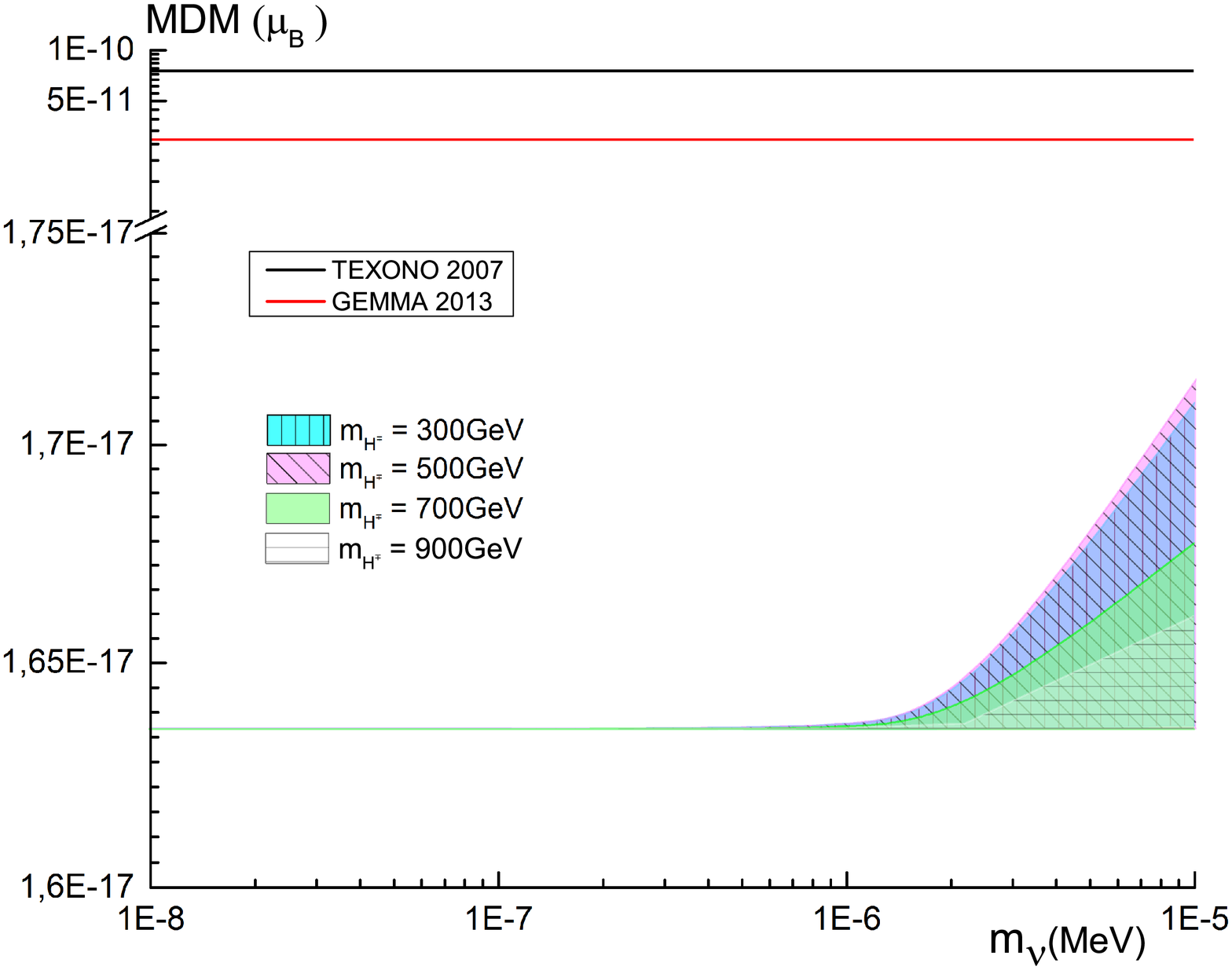} \vspace{0.3cm}
\caption{The graphics show the values of the magnetic dipole moment (MDM) as
a function of the electron neutrino mass between $\left( 1\times
10^{-8}-1\times 10^{-5}\right) MeV$ and for values of the charged Higgs mass
of $\left( 100-300-500-700-900\right) GeV$ for the 2HDM type I (left-hand
side) and $\left( 300-500-700-900\right) GeV$ for the 2HDM type II
(right-hand side).}
\label{fig:el}
\end{figure}

In Fig. \ref{fig:el}, we plot the electron neutrino mass versus MDM for
charged Higgs masses of $100,\ 300,\ 500,\ 700,900\ GeV$ for the 2HDM type I
(left-hand side) and for masses of $300,500,700,900$ $GeV$ for the 2HDM type
II (right-hand side). The horizontal lines correspond to the experimental
upper limits for MDM coming from TEXONO 2007 (Taiwan EXperiment On NeutriNO) 
\cite{Wong} which is $\mu _{\overline{\nu }_{e}}<7.4\times 10^{-11}\mu _{B}$
at $90\%~C.L.$, and GEMMA 2013. (Germanium Experiment for measurement of
Magnetic Moment of Antineutrino)\cite{Gemma} which is $\mu _{\nu
_{e}}<2.9\times 10^{-11}\mu _{B}$ at $90\%~C.L.$. We observe that the
maximum values of MDM that can be reached\ are $1.635\times 10^{-17}\mu _{B}$
for a value of the charged Higgs mass of $m_{H^{\pm }}=100GeV$ and $\tan
\beta =2$ in the case of the 2HDM type I with an electron neutrino mass of $%
1\times 10^{-5}MeV$ and $1.715\times 10^{-17}\mu _{B}$ for a value of the
charged Higgs mass of $m_{H^{\pm }}=500GeV$ and $\tan \beta =69$ in the case
of the 2HDM type II with an electron neutrino mass of $1\times 10^{-5}MeV$ .
These values are far from the experimental threshold and provide no bounds
on the neutrino mass.

\begin{itemize}
\item Muon neutrino case
\end{itemize}

We shall plot within the interval $2\times 10^{-6}MeV\leq m_{\nu _{\mu
}}\leq 4\times 10^{-1}MeV$. Using such an interval in the 2HDM, for
different values of Higgs mass and $\tan \beta $, we can obtain exclusion
regions by taking as reference the experimental thresholds for the MDM of
the muon neutrino. In that way we obtain upper limits for the muon neutrino
mass in this scenario.

\begin{figure}[!htbp]
\centering
\includegraphics[width=0.49\columnwidth]{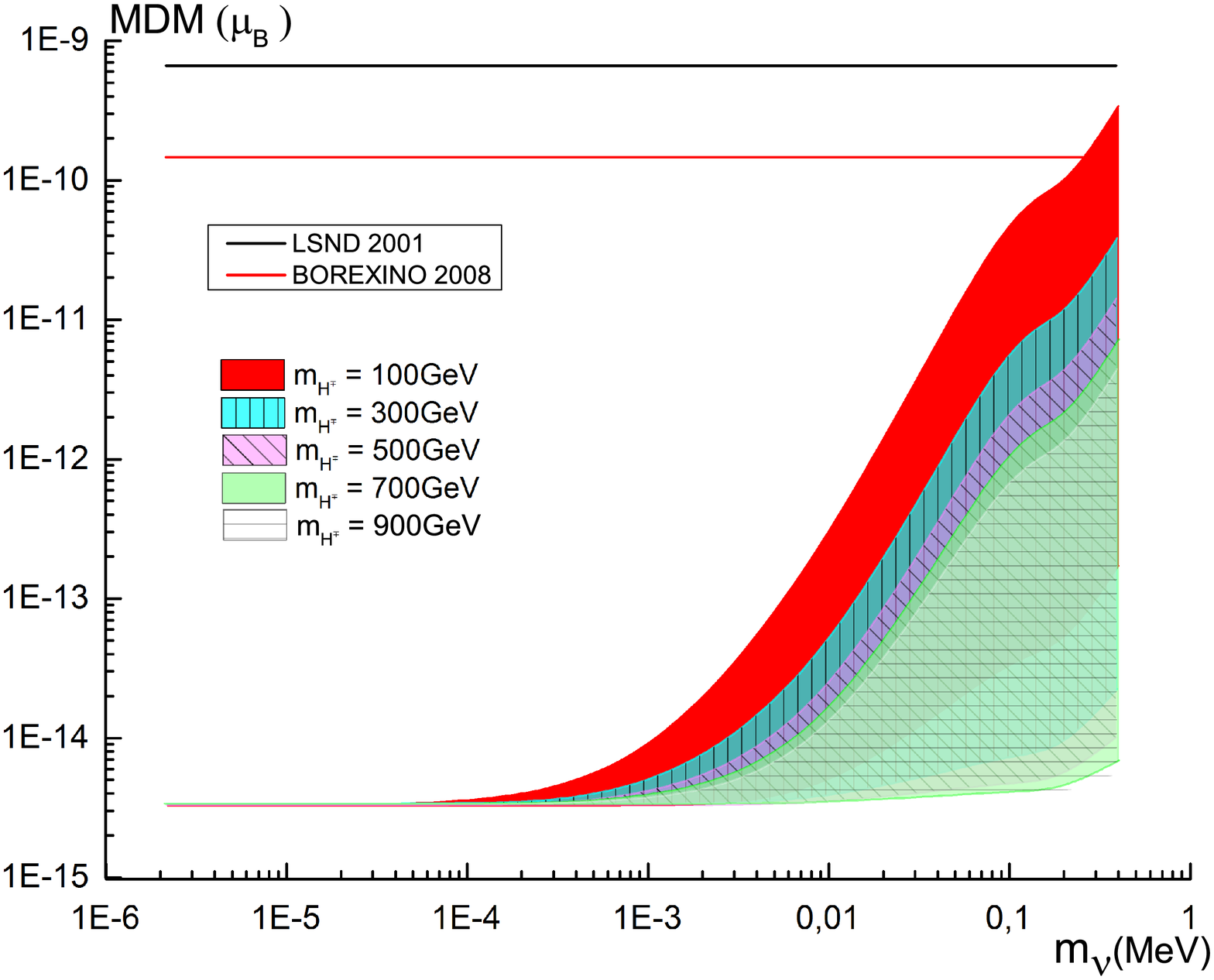} %
\includegraphics[width=0.49\columnwidth]{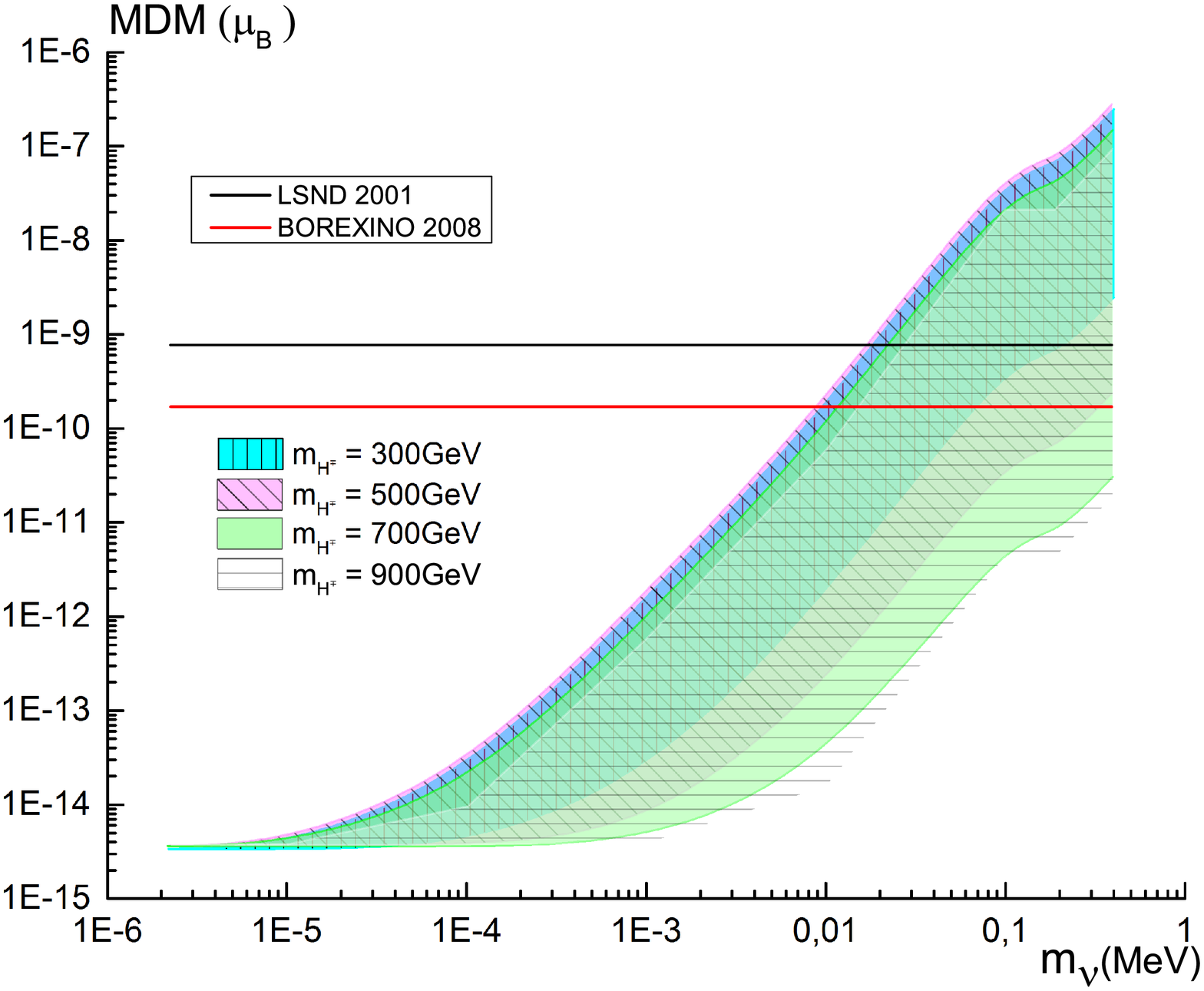} \vspace{0.3cm}
\caption{Values of the magnetic dipole moment (MDM) as a function of the
muon neutrino mass between $\left( 1\times 10^{-5}-4\times 10^{-1}\right) MeV
$ and for values of the charged Higgs mass of $\left(
100-300-500-700-900\right) GeV$ for the 2HDM type I and $\left(
300-500-700-900\right) GeV$ for the 2HDM type II.}
\label{fig:mu}
\end{figure}
\begin{figure}[!htbp]
\centering
\includegraphics[width=0.49\columnwidth]{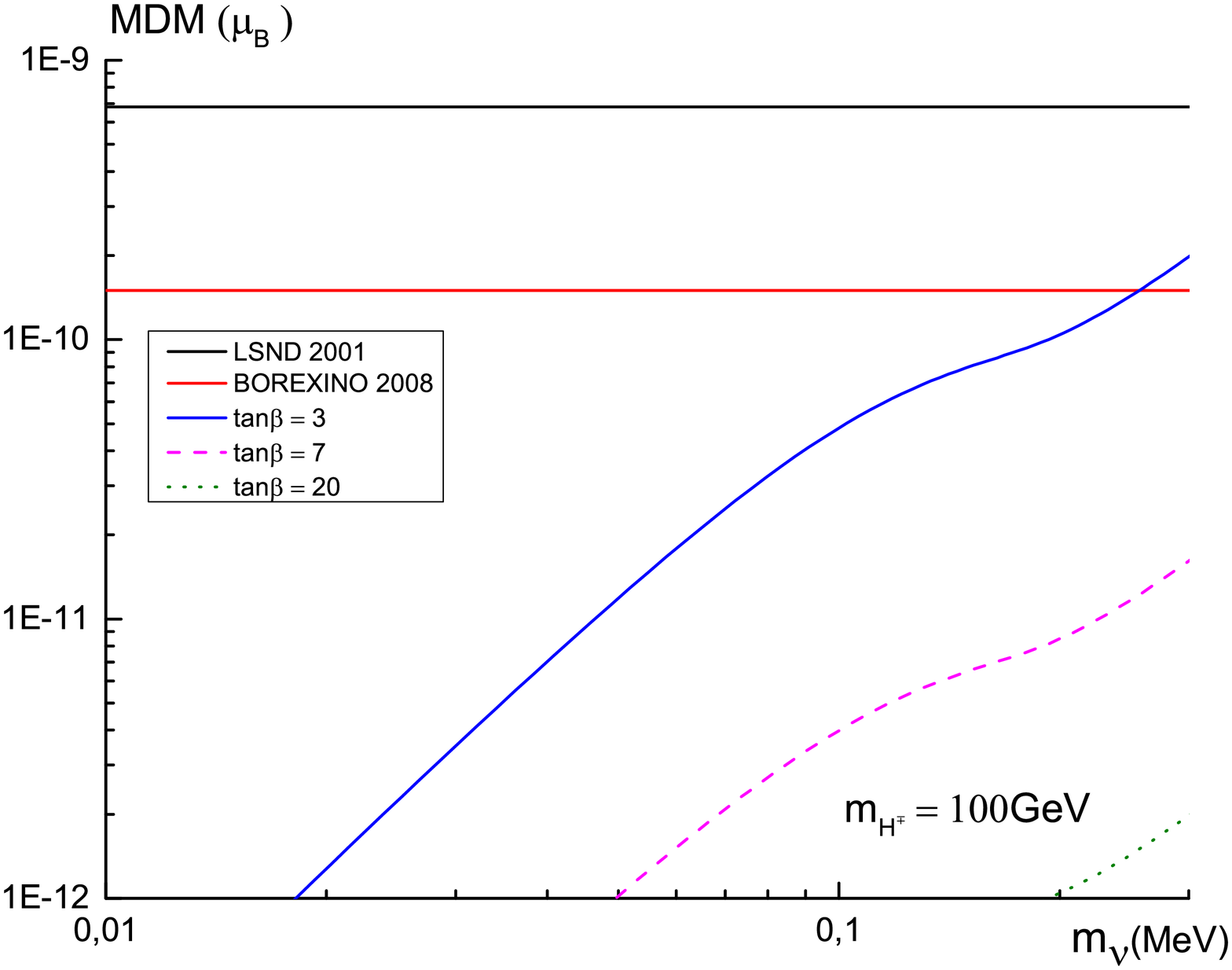} \vspace{0.3cm} 
\caption{Values of the MDM as a function of the muon neutrino mass between $%
\left( 1\times 10^{-2}-4\times 10^{-1}\right) MeV$ and for values of the
charged Higgs mass of $100GeV$ for the 2HDM type I.}
\label{fig:mu_I}
\end{figure}
\begin{figure}[!htbp]
\centering
\includegraphics[width=0.49\columnwidth]{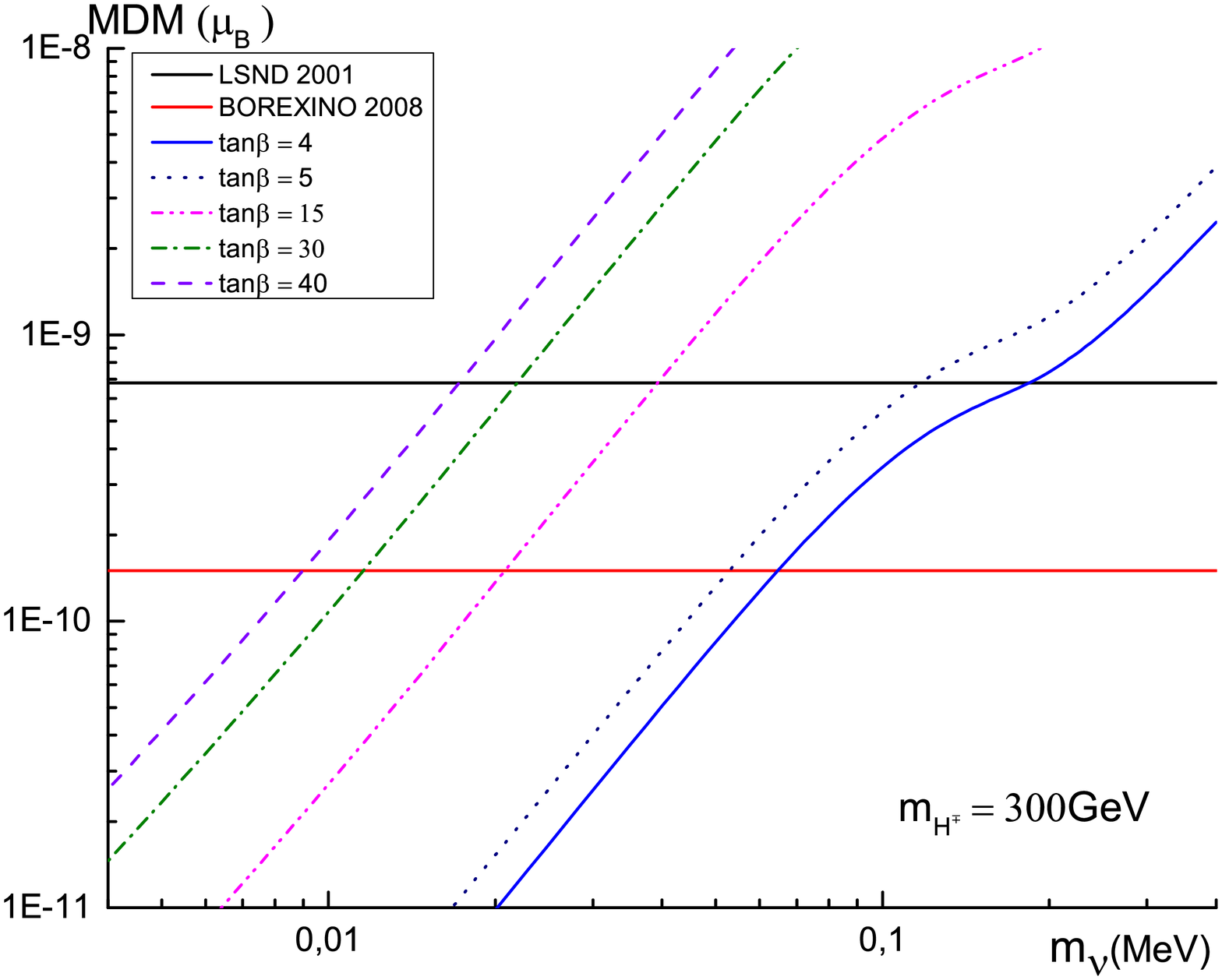} %
\includegraphics[width=0.49\columnwidth]{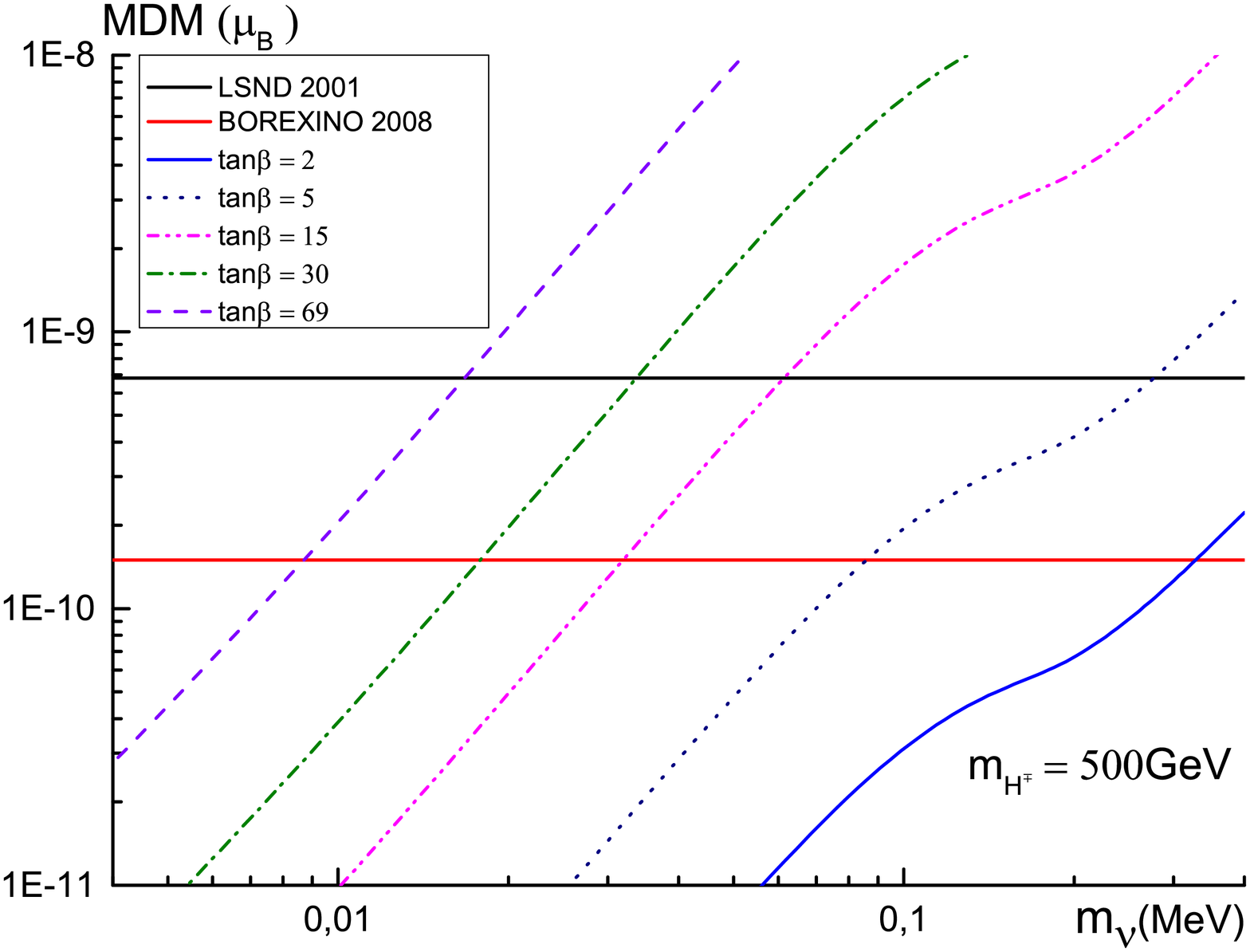} \vspace{0.3cm} %
\includegraphics[width=0.49\columnwidth]{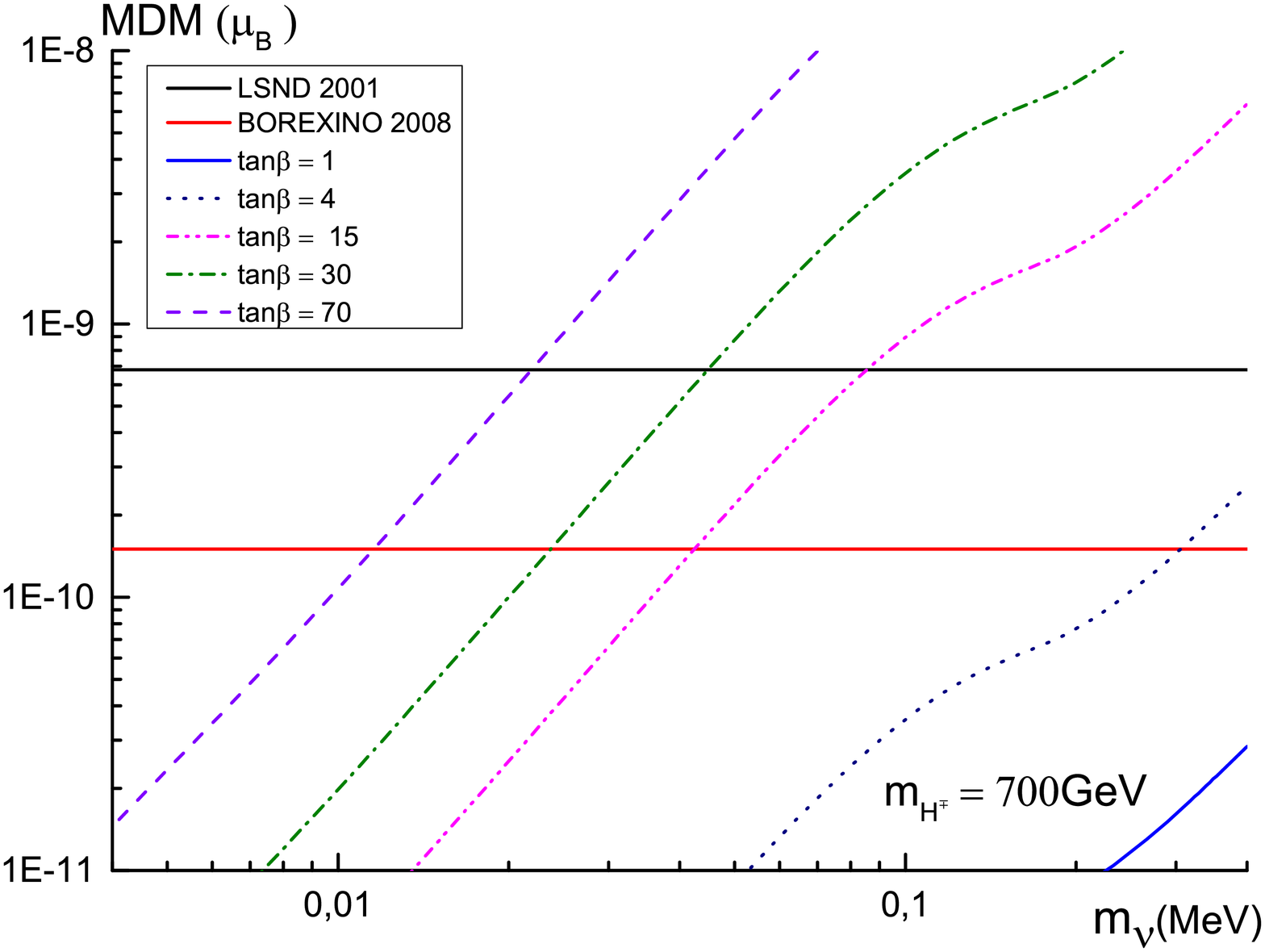} %
\includegraphics[width=0.49\columnwidth]{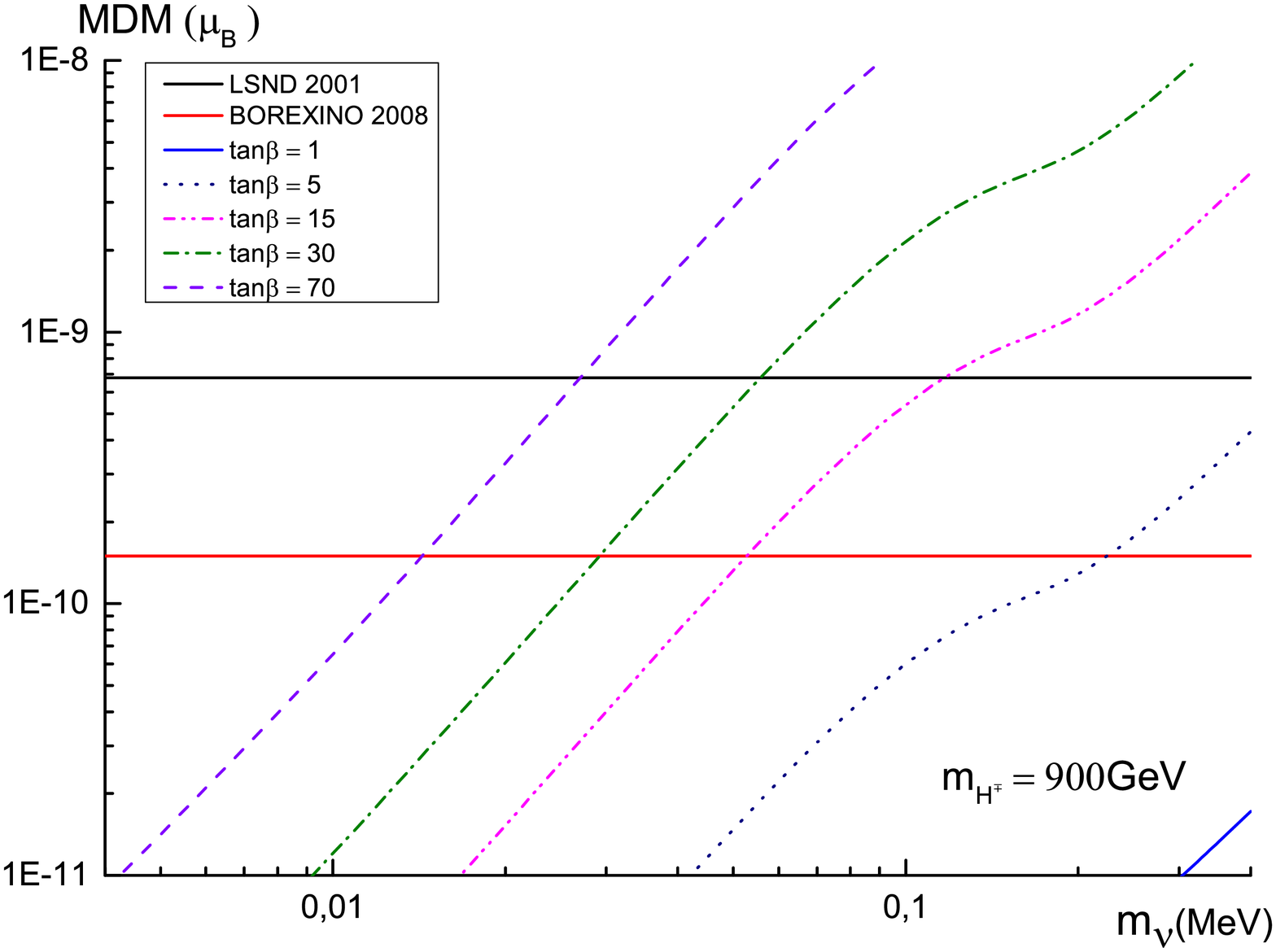} \vspace{0.3cm}
\caption{Values of the MDM as a function of the muon neutrino mass between $%
\left( 1\times 10^{-3}-4\times 10^{-1}\right) MeV$ and for values of the
charged Higgs mass of $\left( 300-500-700-900\right) GeV$ to 2HDM type II}
\label{fig:mu_II}
\end{figure}

\begin{table}[!htbp]
\centering%
\begin{tabular}{|l|c|c|c|c|l|}
\hline
& \multicolumn{4}{c|}{$m_{H^{\pm }}\left( GeV\right) $} &  \\ \cline{2-5}
$\tan \beta $ & $300$ & $500$ & $700$ & $900$ & Experiment \\ \hline
\multirow{2}{1cm}{2} & \multicolumn{1}{|l|}{$-$} & \multicolumn{1}{|l|}{$-$}
& \multicolumn{1}{|l|}{$-$} & \multicolumn{1}{|l|}{$-$} & LSND 2001 \\ 
\cline{2-6}
& \multicolumn{1}{|l|}{$-$} & \multicolumn{1}{|l|}{$3.31\times 10^{-1}$} & 
\multicolumn{1}{|l|}{$-$} & \multicolumn{1}{|l|}{$-$} & BOREXINO 2008 \\ 
\hline
\multirow{2}{1cm}{4} & \multicolumn{1}{|l|}{$1.85\times 10^{-1}$} & 
\multicolumn{1}{|l|}{$-$} & \multicolumn{1}{|l|}{$-$} & \multicolumn{1}{|l|}{%
$-$} & LSND 2001 \\ \cline{2-6}
& \multicolumn{1}{|l|}{$6.47\times 10^{-2}$} & \multicolumn{1}{|l|}{$-$} & 
\multicolumn{1}{|l|}{$-$} & \multicolumn{1}{|l|}{$-$} & BOREXINO 2008 \\ 
\hline
\multirow{2}{1cm}{5} & \multicolumn{1}{|l|}{$1.16\times 10^{-1}$} & 
\multicolumn{1}{|l|}{$2.82\times 10^{-1}$} & \multicolumn{1}{|l|}{$-$} & 
\multicolumn{1}{|l|}{$-$} & LSND 2001 \\ \cline{2-6}
& \multicolumn{1}{|l|}{$5.3\times 10^{-2}$} & \multicolumn{1}{|l|}{$%
8.71\times 10^{-2}$} & \multicolumn{1}{|l|}{$2.18\times 10^{-1}$} & 
\multicolumn{1}{|l|}{$2.25\times 10^{-1}$} & BOREXINO 2008 \\ \hline
\multirow{2}{1cm}{15} & \multicolumn{1}{|l|}{$3.95\times 10^{-2}$} & 
\multicolumn{1}{|l|}{$6.16\times 10^{-2}$} & \multicolumn{1}{|l|}{$%
8.64\times 10^{-2}$} & \multicolumn{1}{|l|}{$1.16\times 10^{-1}$} & LSND 2001
\\ \cline{2-6}
& \multicolumn{1}{|l|}{$2.07\times 10^{-2}$} & \multicolumn{1}{|l|}{$%
3.18\times 10^{-2}$} & \multicolumn{1}{|l|}{$4.26\times 10^{-2}$} & 
\multicolumn{1}{|l|}{$5.27\times 10^{-2}$} & BOREXINO 2008 \\ \hline
\multirow{2}{1cm}{30} & \multicolumn{1}{|l|}{$2.2\times 10^{-2}$} & 
\multicolumn{1}{|l|}{$3.36\times 10^{-2}$} & \multicolumn{1}{|l|}{$%
4.51\times 10^{-2}$} & \multicolumn{1}{|l|}{$5.57\times 10^{-2}$} & LSND 2001
\\ \cline{2-6}
& \multicolumn{1}{|l|}{$1.15\times 10^{-2}$} & \multicolumn{1}{|l|}{$%
1.78\times 10^{-2}$} & \multicolumn{1}{|l|}{$2.39\times 10^{-2}$} & 
\multicolumn{1}{|l|}{$2.89\times 10^{-2}$} & BOREXINO 2008 \\ \hline
\multirow{2}{1cm}{40} & \multicolumn{1}{|l|}{$1.72\times 10^{-2}$} & 
\multicolumn{1}{|l|}{$-$} & \multicolumn{1}{|l|}{$-$} & \multicolumn{1}{|l|}{%
$-$} & LSND 2001 \\ \cline{2-6}
& \multicolumn{1}{|l|}{$8.95\times 10^{-3}$} & \multicolumn{1}{|l|}{$-$} & 
\multicolumn{1}{|l|}{$-$} & \multicolumn{1}{|l|}{$-$} & BOREXINO 2008 \\ 
\hline
\multirow{2}{1cm}{69} & \multicolumn{1}{|l|}{$-$} & \multicolumn{1}{|l|}{$%
1.67\times 10^{-2}$} & \multicolumn{1}{|l|}{$-$} & \multicolumn{1}{|l|}{$-$}
& LSND 2001 \\ \cline{2-6}
& \multicolumn{1}{|l|}{$-$} & \multicolumn{1}{|l|}{$8.7\times 10^{-3}$} & 
\multicolumn{1}{|l|}{$-$} & \multicolumn{1}{|l|}{$-$} & BOREXINO 2008 \\ 
\hline
\multirow{2}{1cm}{70} & \multicolumn{1}{|l|}{$-$} & \multicolumn{1}{|l|}{$-$}
& \multicolumn{1}{|l|}{$2.19\times 10^{-2}$} & \multicolumn{1}{|l|}{$%
2.69\times 10^{-2}$} & LSND 2001 \\ \cline{2-6}
& \multicolumn{1}{|l|}{$-$} & \multicolumn{1}{|l|}{$-$} & 
\multicolumn{1}{|l|}{$1.15\times 10^{-2}$} & \multicolumn{1}{|l|}{$%
1.43\times 10^{-2}$} & BOREXINO 2008 \\ \hline
\end{tabular}%
\vspace{0.4cm}
\caption{This table shows upper bounds for the muon neutrino mass $\left(
MeV\right) $ as a function the free parameters $\tan \protect\beta $ and $%
m_{H^{\pm }}$\ in the 2HDM type II, taken from figure \protect\ref{fig:mu_II}%
. The empty cases correspond to excluded regions of the model.}
\label{tab:muon}
\end{table}

In Fig. \ref{fig:mu}, we plot the muon neutrino mass versus MDM for the same
charged Higgs masses as before for the 2HDM type I (left-hand side) and type
II (right-hand side). The horizontal lines correspond to the experimental
limits for MDM coming from LSND 2001(Liquid Scintillating Neutrino Detector)%
\cite{Auerbach} which is $\mu _{\nu _{\mu }}<6.8\times 10^{-10}\mu _{B}$ at $%
90\%~C.L.$, and BOREXino 2008 (Boron solar neutrino experiment)\cite%
{Montanino} which is $\mu _{\nu _{\mu }}<1.9\times 10^{-10}\mu _{B}$ at $%
90\%~C.L.$.

From Fig. \ref{fig:mu}, we can see that it is necessary to make a more
detailed analysis for certain values of the Higgs mass and the respective
values of $\tan \beta $ for each model, in order to find upper limits for
the neutrino mass as a function of charged Higgs masses, $\tan \beta $ and
the current experimental limits. This more detailed analysis is shown in
Fig. \ref{fig:mu_I} for the 2HDM type I and in Fig. \ref{fig:mu_II} for the
2HDM type II. We observe that in the 2HDM type I the strongest bound for the
muon neutrino mass that we obtain is given by$~2.583\times 10^{-1}MeV\ $for $%
M_{H^{+}}=100GeV\ $and $\tan \beta =3$ obtained from the bound of MDM from
BOREXino. Nevertheless, for the interval of neutrino mass plotted we do not
obtain bounds on the neutrino mass from the bound of MDM coming from LSND,
neither for other masses of the charged Higgs.

As for the 2HDM type II, we observe significant differences in the patterns
of the bounds. For instance, in the 2HDM type II the hierarchy of the bounds
are in opposite order as a function of $\tan \beta $ with respect to the
2HDM type I. This happens because in the 2HDM type I the couplings are
proportional only to $\tan \beta $, while for the 2HDM type II the couplings
are proportional to either\ $\tan \beta $ or $\cot \beta $. Of course, when
the charged Higgs mass increases, the bounds on the muon neutrino mass
become less restrictive since the contribution of the new physics tends to
decouple as the Higgs mass increases.

Considering the current experimental limits, the strongest upper limit that
could take the muon neutrino mass in the 2HDM type II is $8.953\times
10^{-3}MeV,\ $obtained from the set of parameters $m_{H^{\pm }}=300GeV$ and $%
\tan \beta =40$. The numerical values of the upper bounds for the muon
neutrino mass within the 2HDM type II, are shown in table \ref{tab:muon} for
different allowed values of $m_{H^{\pm }}$ and $\tan \beta $ parameters. 
\begin{figure}[!htbp]
\centering
\includegraphics[width=0.49\columnwidth]{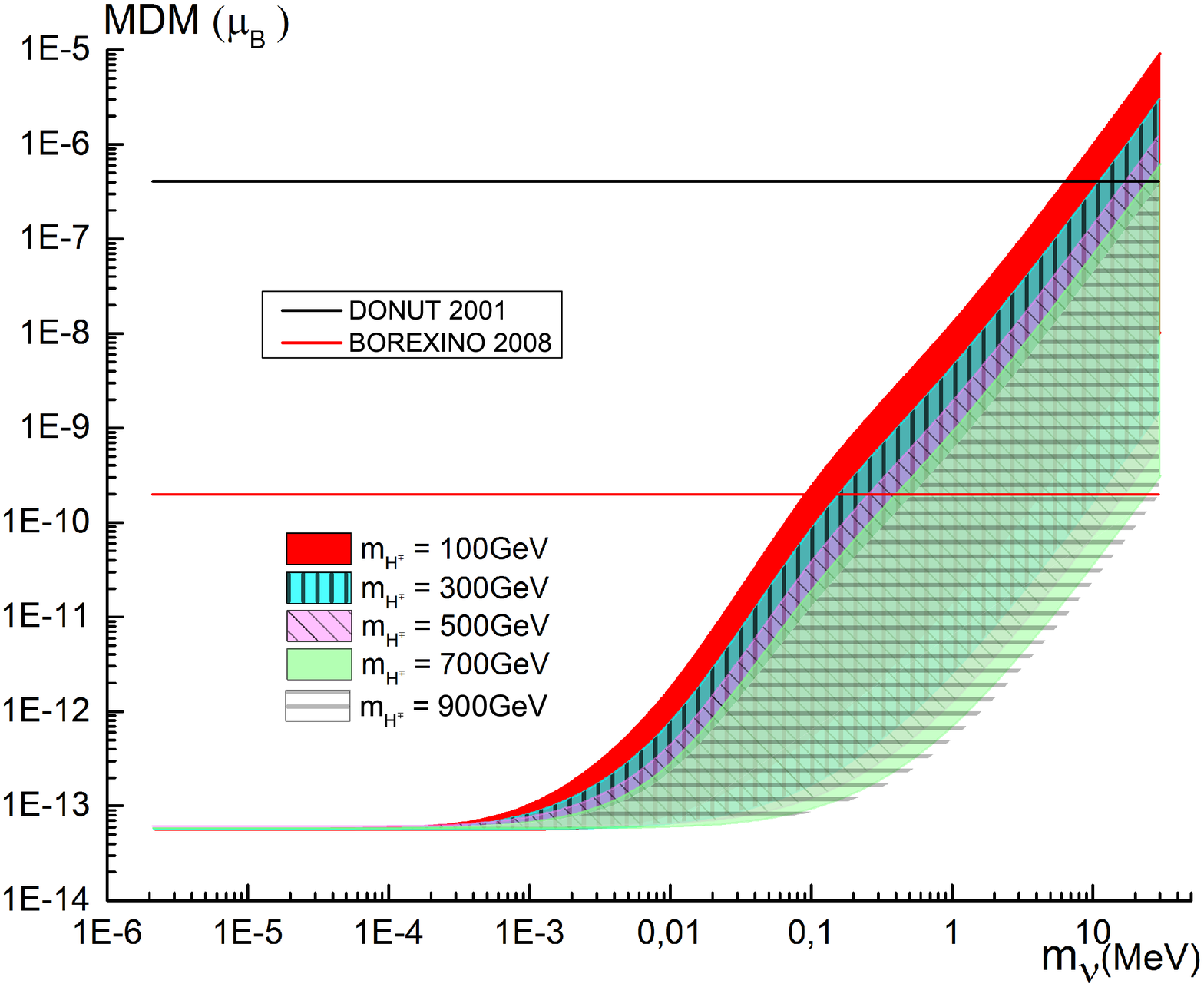} %
\includegraphics[width=0.49\columnwidth]{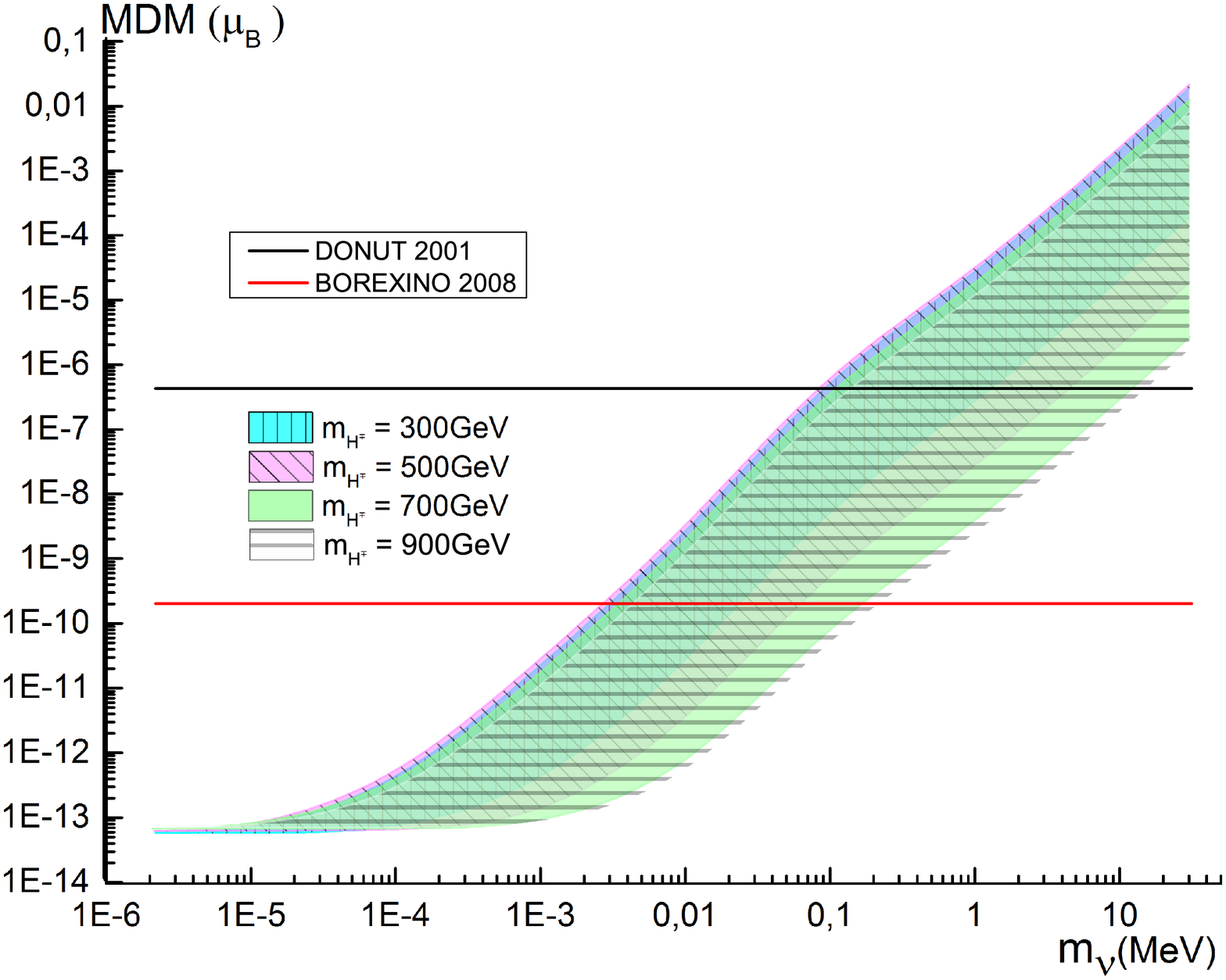} \vspace{0.3cm}
\caption{Values of the MDM as a function of the tau neutrino mass between $%
\left( 1\times 10^{-6}-3\times 10^{1}\right) MeV$ and for values of the
charged Higgs mass of $\left( 100-300-500-700-900\right) GeV$ for the 2HDM
type I and $\left( 300-500-700-900\right) GeV$ for the 2HDM type II.}
\label{fig:ta}
\end{figure}
\begin{figure}[!htbp]
\centering
\includegraphics[width=0.4\columnwidth]{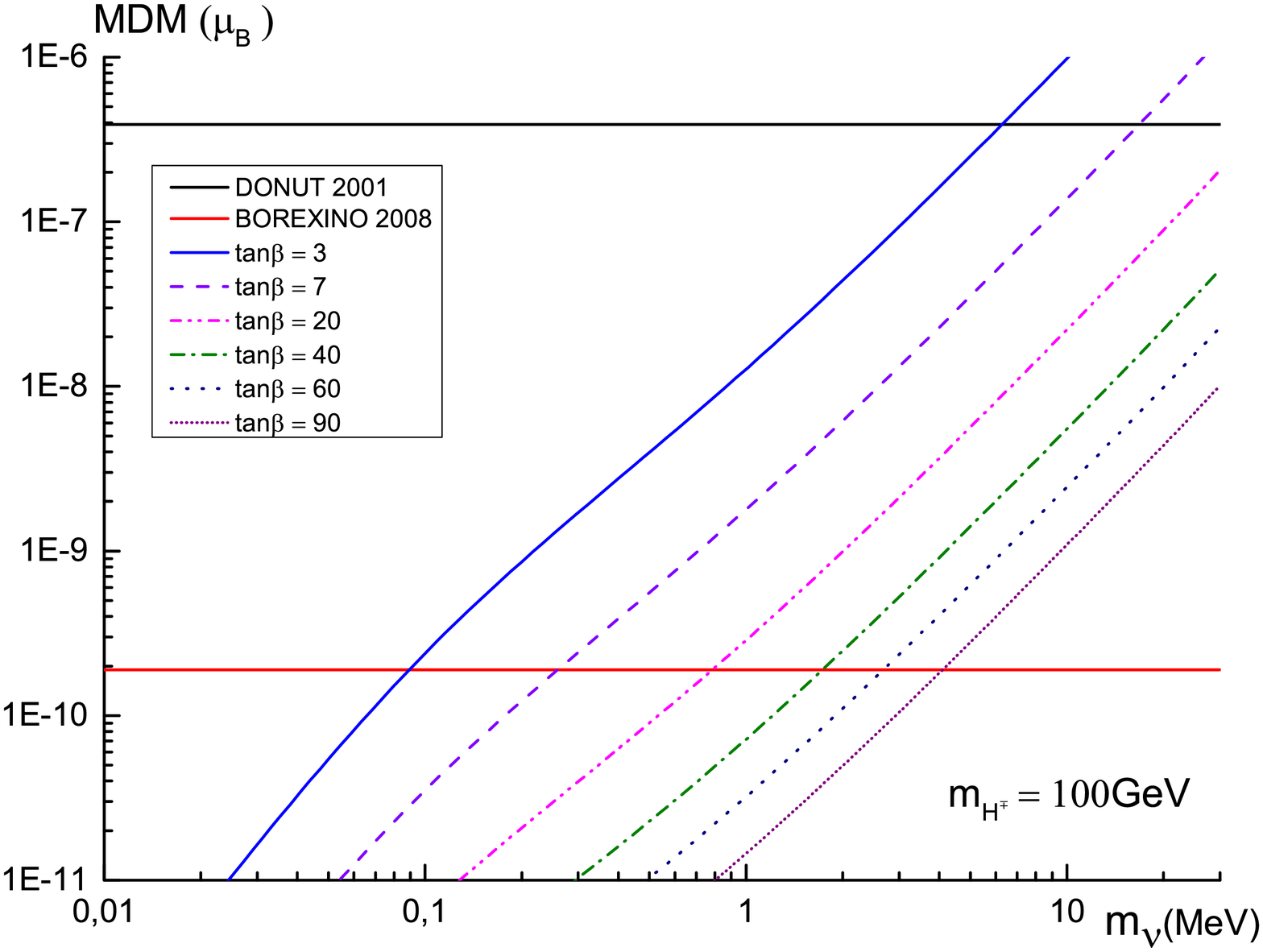} %
\includegraphics[width=0.4\columnwidth]{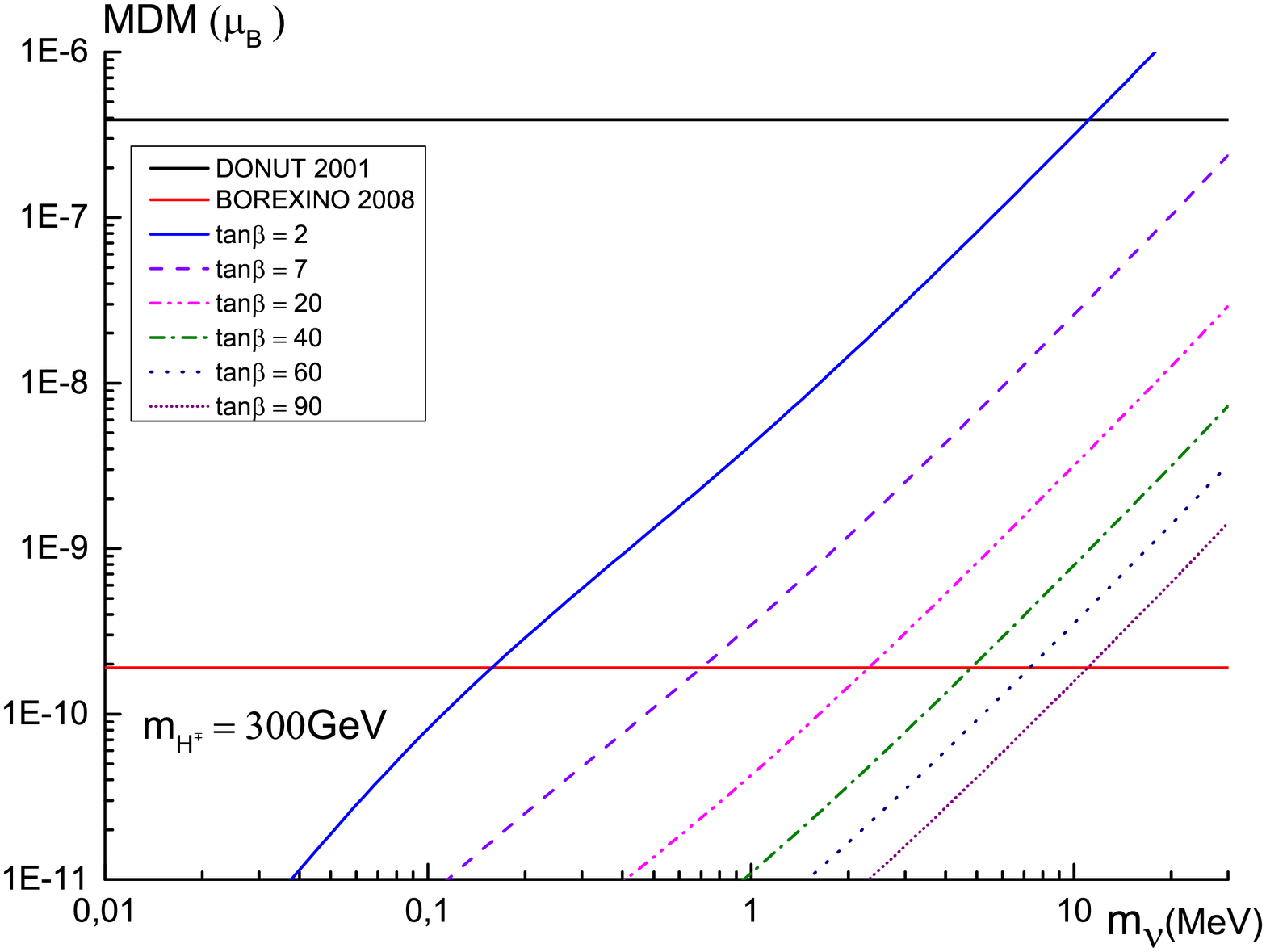} \vspace{0.3cm} %
\includegraphics[width=0.4\columnwidth]{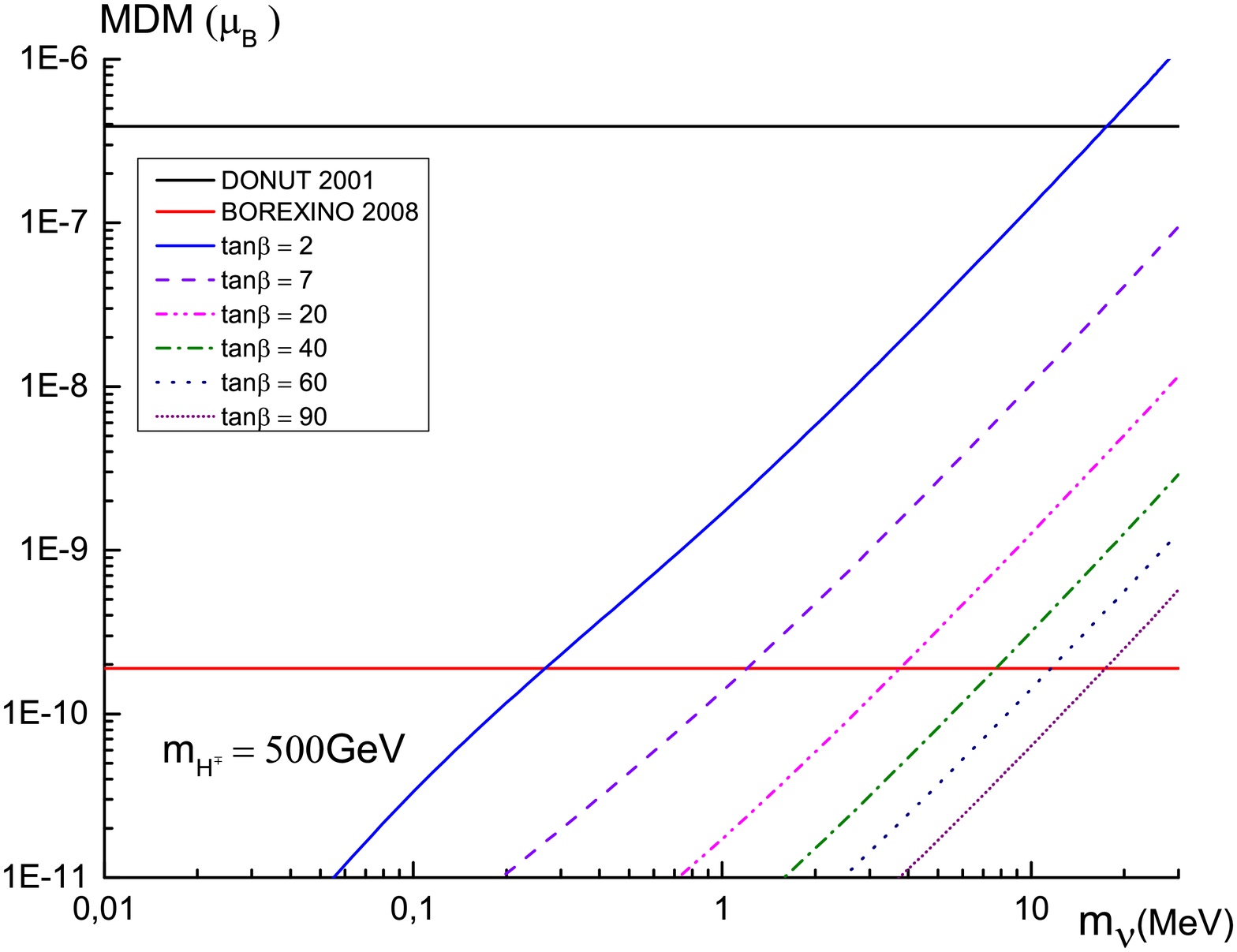} %
\includegraphics[width=0.4\columnwidth]{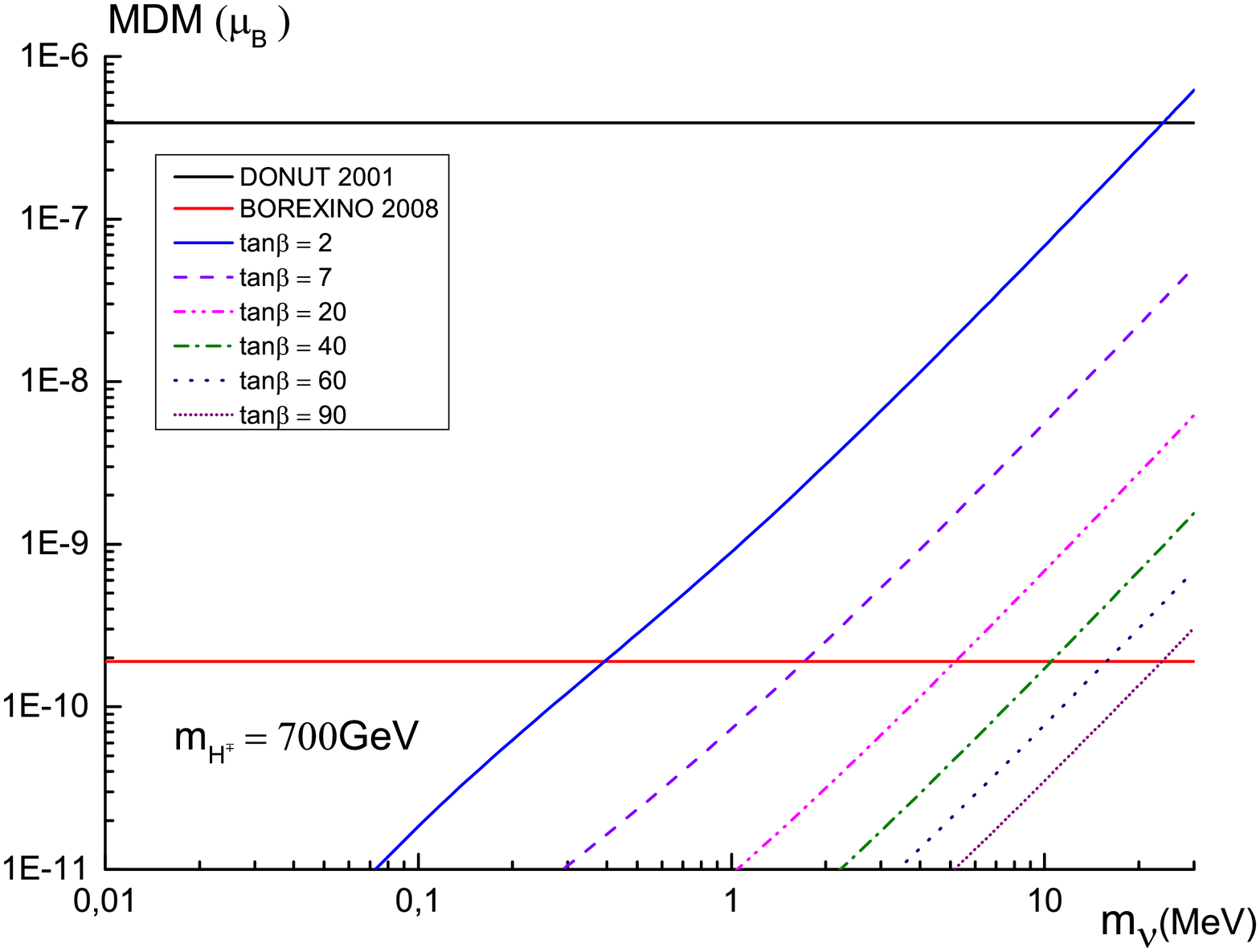} \vspace{0.3cm} %
\includegraphics[width=0.4\columnwidth]{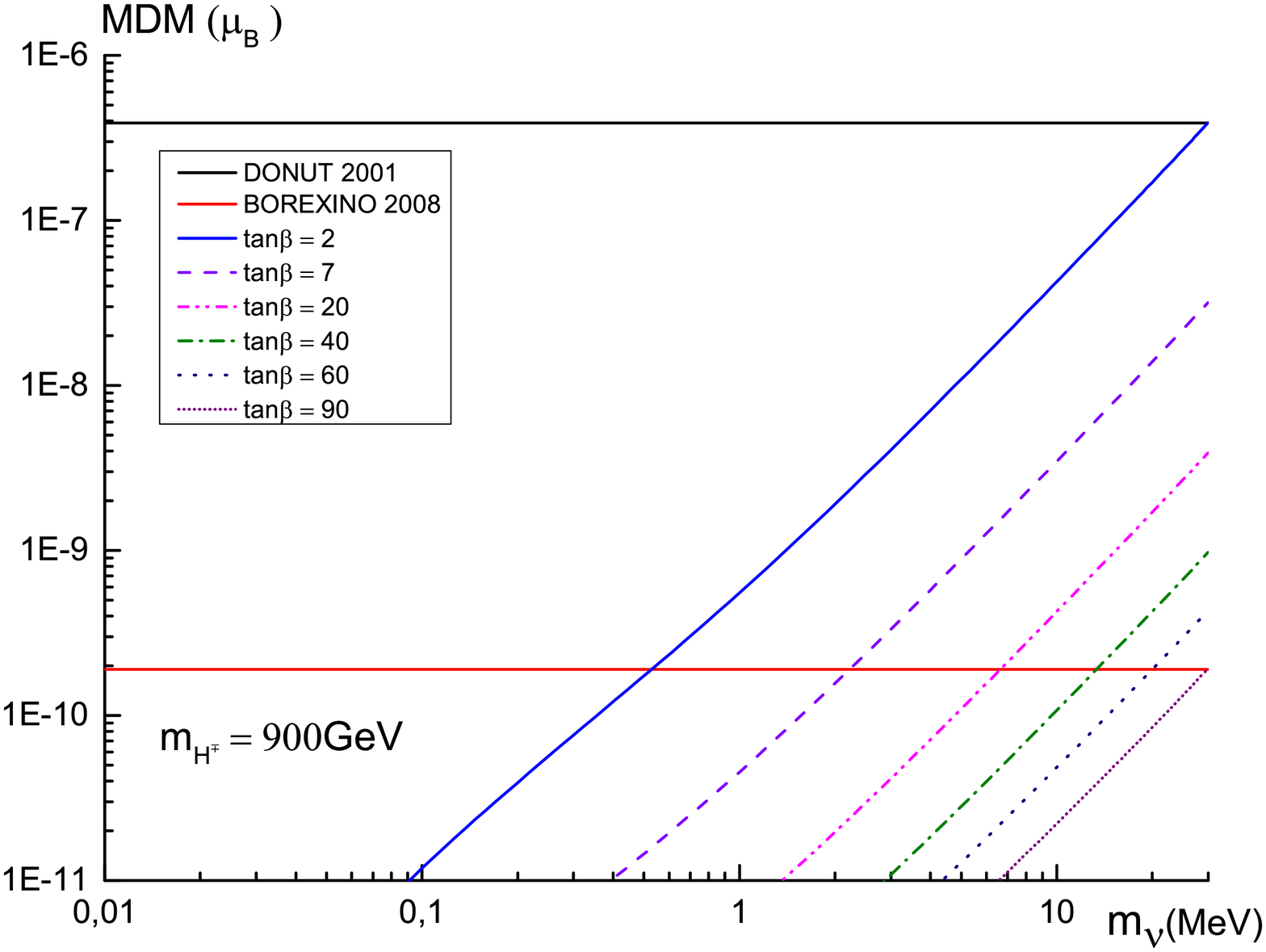} \vspace{0.3cm}
\caption{Values of the MDM as a function of the tau neutrino mass between $%
\left( 1\times 10^{-2}-3\times 10^{1}\right) MeV$ and for values of the
charged Higgs mass of $\left( 100-300-500-700-900\right) GeV$ for the 2HDM
type I.}
\label{fig:ta_I}
\end{figure}

\begin{figure}[!htbp]
\centering
\includegraphics[width=0.49\columnwidth]{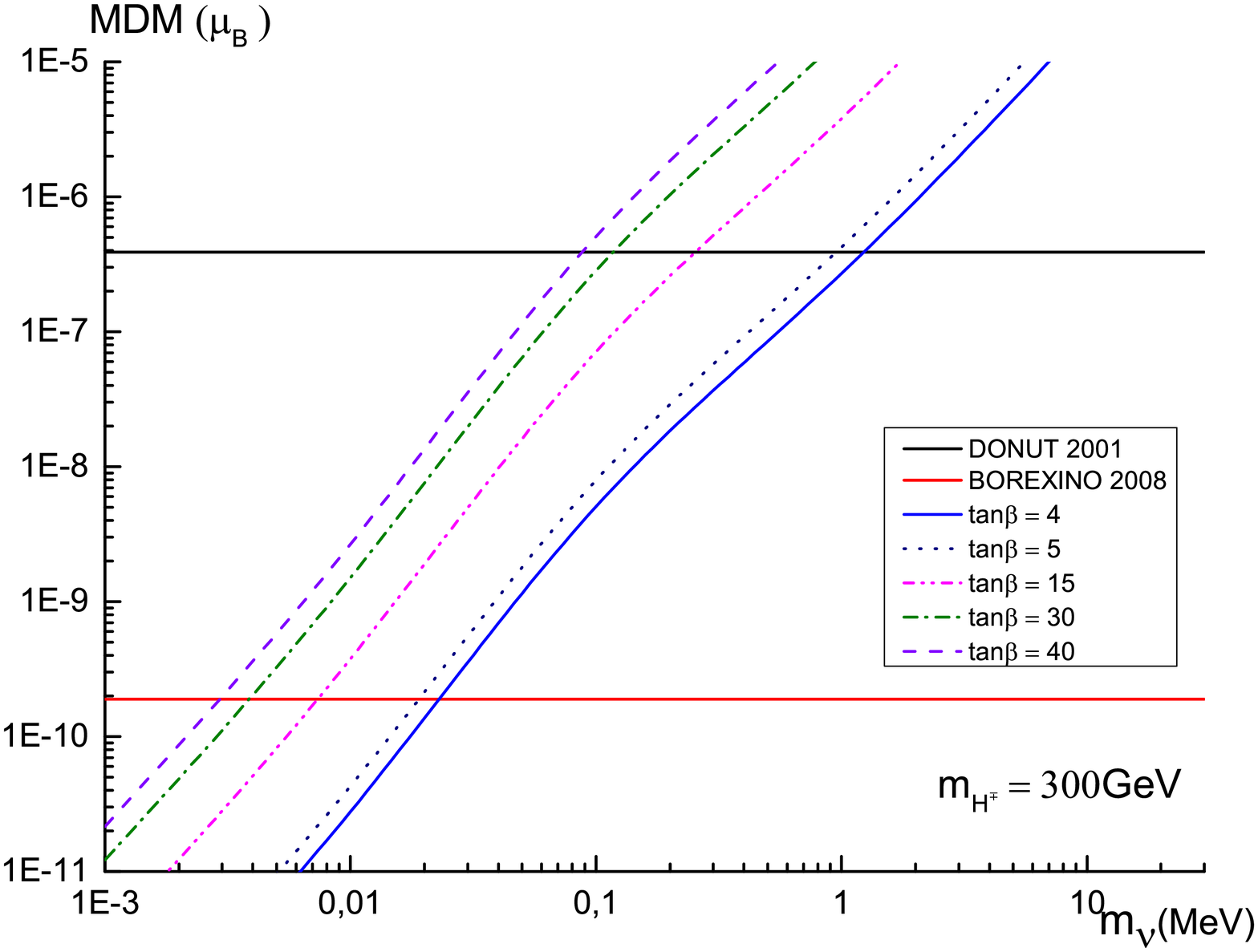} %
\includegraphics[width=0.49\columnwidth]{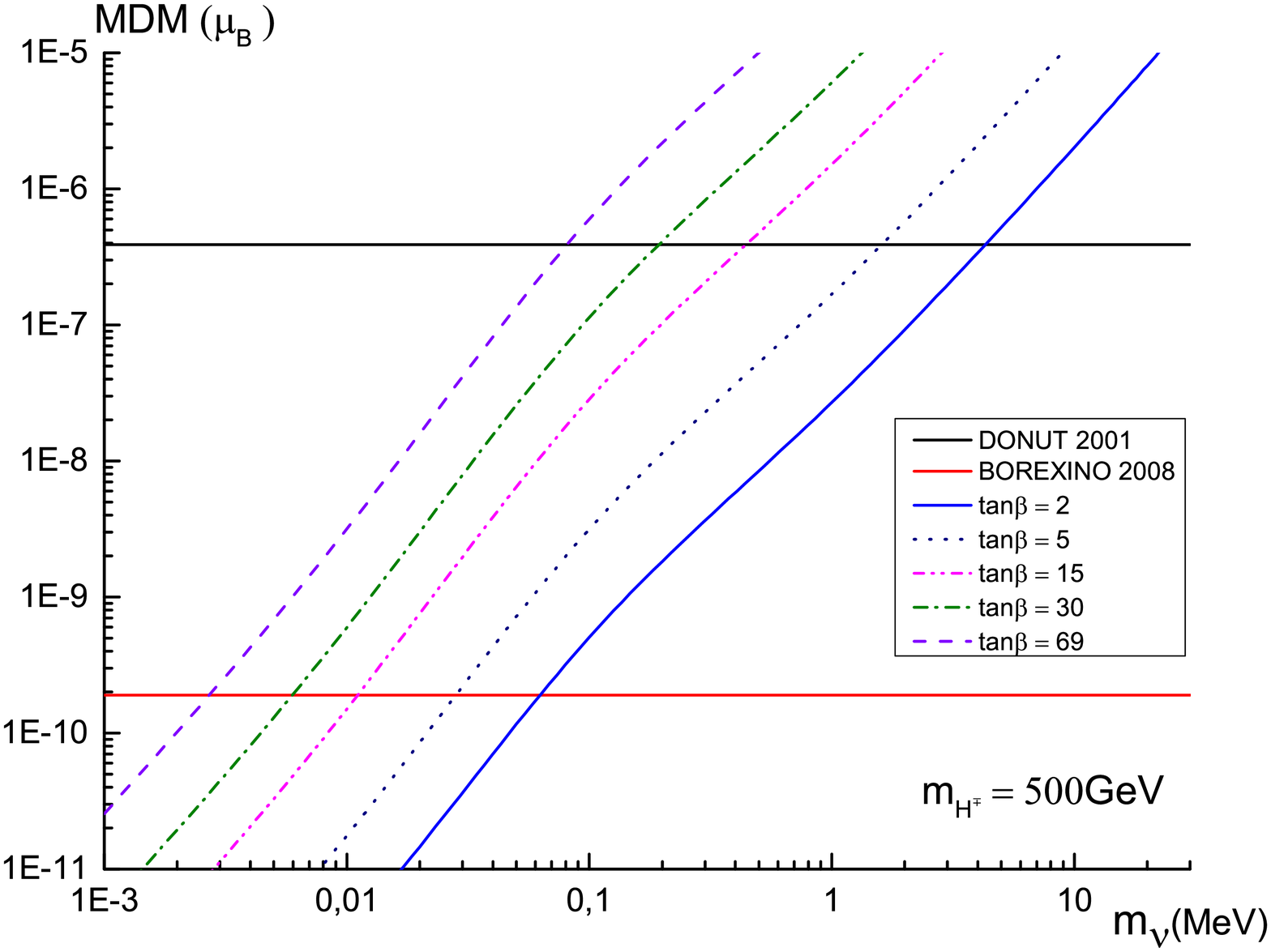} \vspace{0.3cm} %
\includegraphics[width=0.49\columnwidth]{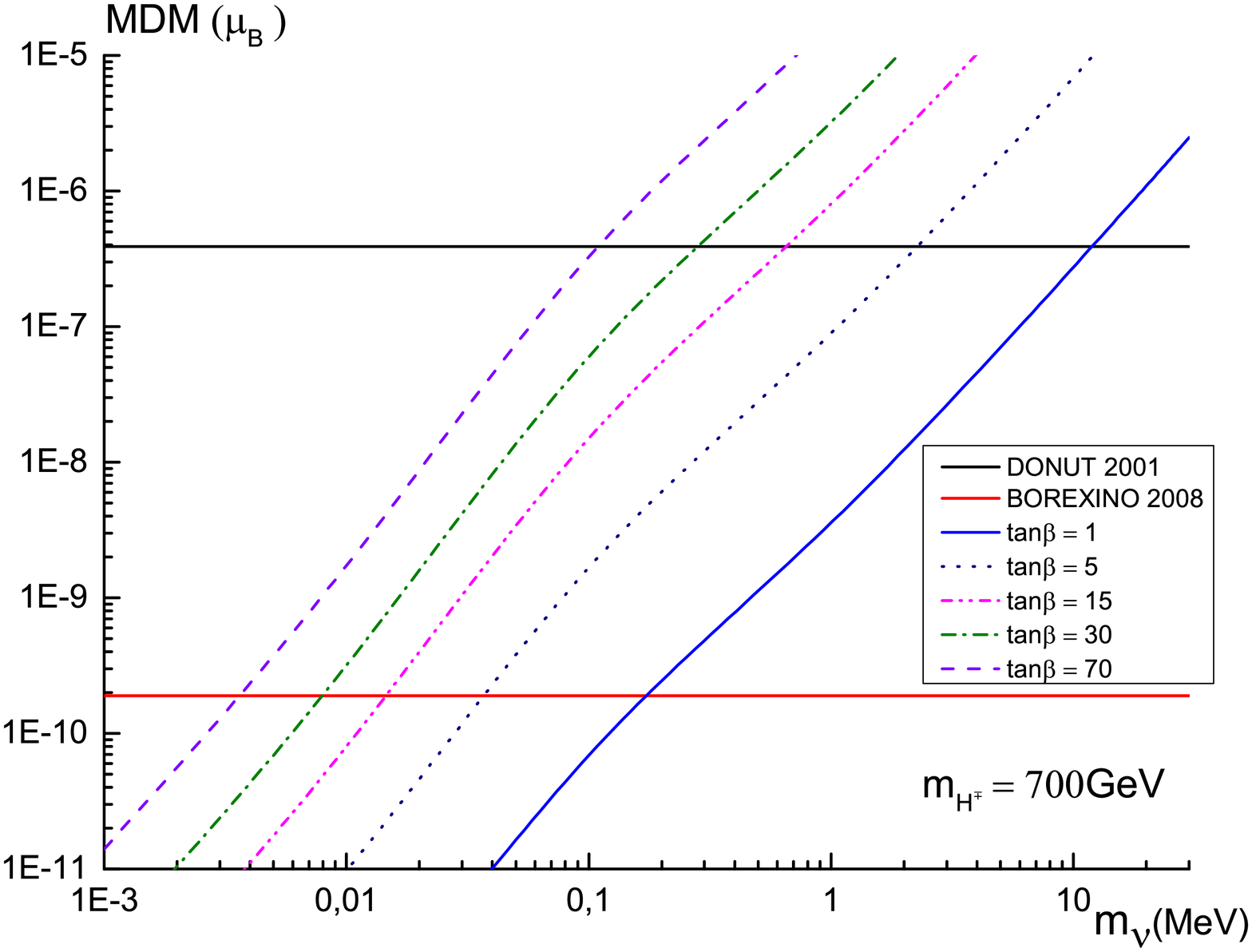} %
\includegraphics[width=0.49\columnwidth]{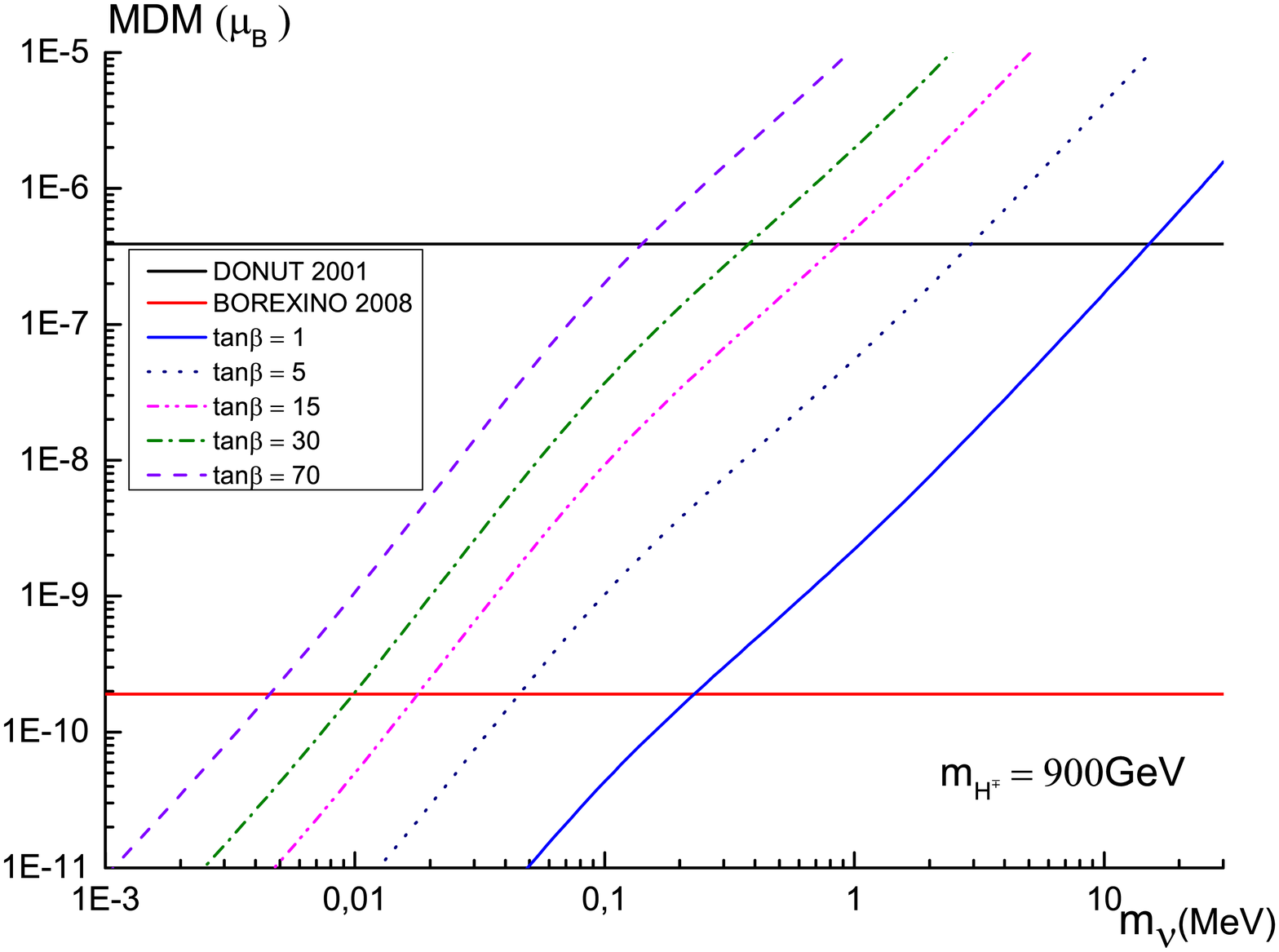} \vspace{0.3cm}
\caption{Values of the MDM as a function of the tau neutrino mass between $%
\left( 1\times 10^{-3}-3\times 10^{1}\right) MeV$ and for values of the
charged Higgs mass of $\left( 300-500-700-900\right) GeV$ for the 2HDM type
II}
\label{fig:ta_II}
\end{figure}
\begin{table}[!htbp]
\centering%
\begin{tabular}{|l|c|c|c|c|c|l|}
\hline
& \multicolumn{5}{c|}{$m_{H^{\pm }}\left( GeV\right) $} &  \\ \cline{2-6}
$\tan \beta $ & $100$ & $300$ & $500$ & $700$ & $900$ & Experiment \\ 
\hline\hline
\multirow{2}{1cm}{2} & \multicolumn{1}{|l|}{$-$} & \multicolumn{1}{|l|}{$%
10.95$} & \multicolumn{1}{|l|}{$17.42$} & \multicolumn{1}{|l|}{$23.95$} & 
\multicolumn{1}{|l|}{$29.78$} & DONUT 2001 \\ \cline{2-6}
& \multicolumn{1}{|l|}{$-$} & \multicolumn{1}{|l|}{$1.55$} & 
\multicolumn{1}{|l|}{$2.71\times 10^{-1}$} & \multicolumn{1}{|l|}{$%
3.95\times 10^{-1}$} & \multicolumn{1}{|l|}{$5.3\times 10^{-1}$} & BOREXINO
2008 \\ \hline
\multirow{2}{1cm}{3} & \multicolumn{1}{|l|}{$6.28$} & \multicolumn{1}{|l|}{$%
- $} & \multicolumn{1}{|l|}{$-$} & \multicolumn{1}{|l|}{$-$} & 
\multicolumn{1}{|l|}{$-$} & DONUT 2001 \\ \cline{2-6}
& \multicolumn{1}{|l|}{$8.86\times 10^{-2}$} & \multicolumn{1}{|l|}{$-$} & 
\multicolumn{1}{|l|}{$-$} & \multicolumn{1}{|l|}{$-$} & \multicolumn{1}{|l|}{%
$-$} & BOREXINO 2008 \\ \hline
\multirow{2}{1cm}{7} & \multicolumn{1}{|l|}{$16.67$} & \multicolumn{1}{|l|}{$%
-$} & \multicolumn{1}{|l|}{$-$} & \multicolumn{1}{|l|}{$-$} & 
\multicolumn{1}{|l|}{$-$} & DONUT 2001 \\ \cline{2-6}
& \multicolumn{1}{|l|}{$2.6\times 10^{-1}$} & \multicolumn{1}{|l|}{$%
7.08\times 10^{-1}$} & \multicolumn{1}{|l|}{$1.19$} & \multicolumn{1}{|l|}{$%
1.74$} & \multicolumn{1}{|l|}{$2.19$} & BOREXINO 2008 \\ \hline
\multirow{2}{1cm}{20} & \multicolumn{1}{|l|}{$-$} & \multicolumn{1}{|l|}{$-$}
& \multicolumn{1}{|l|}{$-$} & \multicolumn{1}{|l|}{$-$} & 
\multicolumn{1}{|l|}{$-$} & DONUT 2001 \\ \cline{2-6}
& \multicolumn{1}{|l|}{$7.74\times 10^{-1}$} & \multicolumn{1}{|l|}{$2.31$}
& \multicolumn{1}{|l|}{$3.75$} & \multicolumn{1}{|l|}{$5.29$} & 
\multicolumn{1}{|l|}{$6.59$} & BOREXINO 2008 \\ \hline
\multirow{2}{1cm}{40} & \multicolumn{1}{|l|}{$-$} & \multicolumn{1}{|l|}{$-$}
& \multicolumn{1}{|l|}{$-$} & \multicolumn{1}{|l|}{$-$} & 
\multicolumn{1}{|l|}{$-$} & DONUT 2001 \\ \cline{2-6}
& \multicolumn{1}{|l|}{$1.72$} & \multicolumn{1}{|l|}{$4.84$} & 
\multicolumn{1}{|l|}{$7.69$} & \multicolumn{1}{|l|}{$10.4$} & 
\multicolumn{1}{|l|}{$13.07$} & BOREXINO 2008 \\ \hline
\multirow{2}{1cm}{60} & \multicolumn{1}{|l|}{$-$} & \multicolumn{1}{|l|}{$-$}
& \multicolumn{1}{|l|}{$-$} & \multicolumn{1}{|l|}{$-$} & 
\multicolumn{1}{|l|}{$-$} & DONUT 2001 \\ \cline{2-6}
& \multicolumn{1}{|l|}{$2.67$} & \multicolumn{1}{|l|}{$7.19$} & 
\multicolumn{1}{|l|}{$11.81$} & \multicolumn{1}{|l|}{$15.68$} & 
\multicolumn{1}{|l|}{$19.35$} & BOREXINO 2008 \\ \hline
\multirow{2}{1cm}{90} & \multicolumn{1}{|l|}{$-$} & \multicolumn{1}{|l|}{$-$}
& \multicolumn{1}{|l|}{$-$} & \multicolumn{1}{|l|}{$-$} & 
\multicolumn{1}{|l|}{$-$} & DONUT 2001 \\ \cline{2-6}
& \multicolumn{1}{|l|}{$4.05$} & \multicolumn{1}{|l|}{$10.95$} & 
\multicolumn{1}{|l|}{$17.42$} & \multicolumn{1}{|l|}{$23.64$} & 
\multicolumn{1}{|l|}{$29.4$} & BOREXINO 2008 \\ \hline
\end{tabular}%
\vspace{0.4cm}
\caption{This table shows the upper bounds for the tau neutrino mass $\left(
MeV\right) $ for several allowed values of the free parameters $\tan \protect%
\beta $ and $m_{H^{\pm }}$ in the 2HDM type I, taken from figure \protect\ref%
{fig:ta_I}. The empty cases correspond to excluded regions of the model.}
\label{tab:tau1}
\end{table}

\begin{table}[!htbp]
\centering%
\begin{tabular}{|l|c|c|c|c|l|}
\hline
& \multicolumn{4}{c|}{$m_{H^{\pm }}\left( GeV\right) $} &  \\ \cline{2-5}
$\tan \beta $ & $300$ & $500$ & $700$ & $900$ & Experiment \\ \hline
\multirow{2}{1cm}{1} & \multicolumn{1}{|l|}{$-$} & \multicolumn{1}{|l|}{$-$}
& \multicolumn{1}{|l|}{$11.87$} & \multicolumn{1}{|l|}{$15.65$} & DONUT 2001
\\ \cline{2-6}
& \multicolumn{1}{|l|}{$-$} & \multicolumn{1}{|l|}{$-$} & 
\multicolumn{1}{|l|}{$1.73\times 10^{-1}$} & \multicolumn{1}{|l|}{$%
2.27\times 10^{-1}$} & BOREXINO 2008 \\ \hline
\multirow{2}{1cm}{2} & \multicolumn{1}{|l|}{$-$} & \multicolumn{1}{|l|}{$%
4.44 $} & \multicolumn{1}{|l|}{$-$} & \multicolumn{1}{|l|}{$-$} & DONUT 2001
\\ \cline{2-6}
& \multicolumn{1}{|l|}{$-$} & \multicolumn{1}{|l|}{$6.24\times 10^{-2}$} & 
\multicolumn{1}{|l|}{$-$} & \multicolumn{1}{|l|}{$-$} & BOREXINO 2008 \\ 
\hline
\multirow{2}{1cm}{4} & \multicolumn{1}{|l|}{$1.24$} & \multicolumn{1}{|l|}{$%
- $} & \multicolumn{1}{|l|}{$-$} & \multicolumn{1}{|l|}{$-$} & DONUT 2001 \\ 
\cline{2-6}
& \multicolumn{1}{|l|}{$2.27\times 10^{-2}$} & \multicolumn{1}{|l|}{$-$} & 
\multicolumn{1}{|l|}{$-$} & \multicolumn{1}{|l|}{$-$} & BOREXINO 2008 \\ 
\hline
\multirow{2}{1cm}{5} & \multicolumn{1}{|l|}{$9.52\times 10^{-1}$} & 
\multicolumn{1}{|l|}{$1.58$} & \multicolumn{1}{|l|}{$2.27$} & 
\multicolumn{1}{|l|}{$2.97$} & DONUT 2001 \\ \cline{2-6}
& \multicolumn{1}{|l|}{$1.82\times 10^{-2}$} & \multicolumn{1}{|l|}{$%
2.77\times 10^{-2}$} & \multicolumn{1}{|l|}{$3.66\times 10^{-2}$} & 
\multicolumn{1}{|l|}{$4.52\times 10^{-2}$} & BOREXINO 2008 \\ \hline
\multirow{2}{1cm}{15} & \multicolumn{1}{|l|}{$2.48\times 10^{-1}$} & 
\multicolumn{1}{|l|}{$4.58\times 10^{-1}$} & \multicolumn{1}{|l|}{$%
6.49\times 10^{-1}$} & \multicolumn{1}{|l|}{$8.63\times 10^{-1}$} & DONUT
2001 \\ \cline{2-6}
& \multicolumn{1}{|l|}{$7.13\times 10^{-3}$} & \multicolumn{1}{|l|}{$%
1.13\times 10^{-2}$} & \multicolumn{1}{|l|}{$1.43\times 10^{-2}$} & 
\multicolumn{1}{|l|}{$1.73\times 10^{-2}$} & BOREXINO 2008 \\ \hline
\multirow{2}{1cm}{30} & \multicolumn{1}{|l|}{$1.14\times 10^{-1}$} & 
\multicolumn{1}{|l|}{$1.93\times 10^{-1}$} & \multicolumn{1}{|l|}{$%
2.89\times 10^{-1}$} & \multicolumn{1}{|l|}{$3.82\times 10^{-1}$} & DONUT
2001 \\ \cline{2-6}
& \multicolumn{1}{|l|}{$3.7\times 10^{-3}$} & \multicolumn{1}{|l|}{$%
5.82\times 10^{-3}$} & \multicolumn{1}{|l|}{$8.04\times 10^{-3}$} & 
\multicolumn{1}{|l|}{$9.62\times 10^{-3}$} & BOREXINO 2008 \\ \hline
\multirow{2}{1cm}{40} & \multicolumn{1}{|l|}{$8.58\times 10^{-2}$} & 
\multicolumn{1}{|l|}{$-$} & \multicolumn{1}{|l|}{$-$} & \multicolumn{1}{|l|}{%
$-$} & DONUT 2001 \\ \cline{2-6}
& \multicolumn{1}{|l|}{$2.87\times 10^{-3}$} & \multicolumn{1}{|l|}{$-$} & 
\multicolumn{1}{|l|}{$-$} & \multicolumn{1}{|l|}{$-$} & BOREXINO 2008 \\ 
\hline
\multirow{2}{1cm}{69} & \multicolumn{1}{|l|}{$-$} & \multicolumn{1}{|l|}{$%
8.14\times 10^{-2}$} & \multicolumn{1}{|l|}{$-$} & \multicolumn{1}{|l|}{$-$}
& DONUT 2001 \\ \cline{2-6}
& \multicolumn{1}{|l|}{$-$} & \multicolumn{1}{|l|}{$2.71\times 10^{-3}$} & 
\multicolumn{1}{|l|}{$-$} & \multicolumn{1}{|l|}{$-$} & BOREXINO 2008 \\ 
\hline
\multirow{2}{1cm}{70} & \multicolumn{1}{|l|}{$-$} & \multicolumn{1}{|l|}{$-$}
& \multicolumn{1}{|l|}{$1.12\times 10^{-1}$} & \multicolumn{1}{|l|}{$%
1.39\times 10^{-1}$} & DONUT 2001 \\ \cline{2-6}
& \multicolumn{1}{|l|}{$-$} & \multicolumn{1}{|l|}{$-$} & 
\multicolumn{1}{|l|}{$3.57\times 10^{-3}$} & \multicolumn{1}{|l|}{$%
4.48\times 10^{-3}$} & BOREXINO 2008 \\ \hline
\end{tabular}%
\vspace{0.4cm}
\caption{Upper bounds for the tau neutrino mass $\left( MeV\right) $ for
several allowed values of the free parameters $\tan \protect\beta $ and $%
m_{H^{\pm }}$ in the 2HDM type II, taken from figure \protect\ref{fig:ta_II}%
. The empty cases correspond to excluded regions of the model.}
\label{tab:tau2}
\end{table}

\begin{itemize}
\item Tau neutrino case
\end{itemize}

We shall plot within the interval $2\times 10^{-6}\leq m_{\nu _{\tau }}\leq
20MeV$, and obtain our bounds from the experimental limits on the MDM. In
Fig. \ref{fig:ta}, we plot the tau neutrino mass versus MDM for the same
charged Higgs masses as before for the 2HDM type I (left-hand side) and type
II (right-hand side). The horizontal lines correspond to the experimental
limits for MDM coming from DONUT 2001(Direct Observation of the NU Tau)\cite%
{Schwienhorst} which is $\mu _{\nu _{\tau }}<3.9\times 10^{-7}\mu _{B}$ at $%
90\%~C.L.$, and BOREXino 2008\cite{Montanino} which is $\mu _{\nu _{\mu
}}<1.5\times 10^{-10}\mu _{B}$ at $90\%~C.L$.

Figure \ref{fig:ta} shows that we require a more detailed analysis of the
upper bounds of the tau neutrino masses in terms of the free parameters.
Such an analysis is carried out in Fig. \ref{fig:ta_I} for the 2HDM type I
and in Fig. \ref{fig:ta_II} for the 2HDM type II. Once again we have
significant differences in the patterns of the bounds for the models of type
I and of type II because of the different behavior of the couplings with
respect to the $\tan \beta $ parameter. Further, the upper limits of the tau
neutrino mass is weakened as the charged Higgs mass increases owing to the
decoupling behavior of the diagrams with respect to the Higgs mass.

Considering the current experimental limits, the strongest limit obtained
for the tau neutrino mass in the 2HDM type I is$\ 9.098\times 10^{-2}MeV$,
and occurs for the values $m_{H^{\pm }}=100GeV$ and $\tan \beta =3$. As for
the 2HDM type II the strongest upper limit on the tau neutrino mass is $%
2.926\times 10^{-3}MeV$ obtained with the set of parameters $m_{H^{\pm
}}=300GeV$ and $\tan \beta =40$. Finally, the numerical values of the upper
bounds for the muon neutrino mass within the 2HDM type I and II, are shown
in tables \ref{tab:tau1} and \ref{tab:tau2} respectively, for different
allowed values of $m_{H^{\pm }}$ and $\tan \beta $ parameters.

\section{Conclusions\label{sec:conclusions}}

The neutrino magnetic moment provides a tool for exploration of physics
beyond the Standard Model. The magnitude of the magnetic moment is highly
sensitive to the neutrino mass, but also depends on the mass of the
associated charged lepton inserted into the loops.

Further, the value of the magnetic moments of the neutrinos could be
modified with physics beyond the Standard Model. In particular for the Two
Higgs Doublet Model (2HDM) we evaluated the contributions coming from the
insertion of the charged Higgs boson into the loops. Our results show that
for the 2HDM of type I and of type II, the total contribution is far from
the threshold of experimental detection in the case of electron neutrinos
(owing to the supression coming from the electron mass into the loops),
obtaining a maximum contribution about six orders of magnitude below the
present experimental limits. In the case of muon neutrinos the total
contribution produces weak bounds for the mass of the neutrinos for model
type I, and stronger bounds for the case of model type II. Finally, such
bounds are much stronger for tau neutrinos (because of the enhancement of
the tau mass into the loops) for either type of 2HDM, but restrictions are
much stronger for the model type II.

In general since the bounds are highly sensitive to the value of the $\tan
\beta $ parameter, the limits obtained are significantly different for the
model type I with respect to the model type II because of the different
dependence on the Yukawa couplings of each model with $\tan \beta $. Of
course, the limits are weakened as the mass of the charged Higgs increases
since the contribution of new physics tends to decouple as the Higss mass
grows. Further, since the contribution of diagrams involving the associated
charged lepton increases with the mass of the charged lepton, the strongest
bounds are obtained for the neutrino associated with the heaviest charged
lepton (tau neutrino) while basically no bounds near the experimental
threshold are obtained for the electron neutrino.

\section{Acknowledgments}

We acknowledge to Division de Investigacion de Bogotá (DIB) for its
financial support and the Universidad Manuela Beltran.

{\small \appendix}

\section{Explicit expressions for the EFF}

In this Appendix we present some details of the process of calculating the
EFF's. For the case of two higgs bosons and one lepton $\left( 2H^{\pm
}1L\right) $, the general form of the contribution can be written as 
\begin{equation*}
\Lambda _{2H^{\pm }1L}^{\alpha }\left( q,l\right) =-e\int \frac{d^{4}k}{%
\left( 2\pi \right) ^{4}}\left\{ \frac{\left( 2k^{\alpha }+p_{2}^{\alpha
}+p_{1}^{\alpha }\right) \left( aP_{L}+bP_{R}\right) \left( \NEG%
{k}+m_{l}\right) \left( cP_{L}+dP_{R}\right) }{\left[ \left( k+p_{1}\right)
^{2}-m_{H^{\pm }}^{2}\right] \left[ \left( k+p_{2}\right) ^{2}-m_{H^{\pm
}}^{2}\right] \left( k^{2}-m_{l}^{2}\right) }\right\}
\end{equation*}%
expanding the numerator, denoting $A=\left( aP_{L}+bP_{R}\right) $ and $%
B=\left( cP_{L}+dP_{R}\right) $ we find 
\begin{equation*}
\left( 2k^{\alpha }+p_{2}^{\alpha }+p_{1}^{\alpha }\right) A\left( \NEG%
{k}+m_{l}\right) B=\left( 2k^{\alpha }k_{\beta }+p_{2}^{\alpha }k_{\beta
}+p_{1}^{\alpha }k_{\beta }\right) A\gamma ^{\beta }B+m_{l}\left( 2k^{\alpha
}+p_{2}^{\alpha }+p_{1}^{\alpha }\right) AB
\end{equation*}%
and for the denominator we use the dimensional regularization method%
\begin{equation*}
\frac{1}{\underset{a_{1}}{\underbrace{\left[ \left( k+p_{2}\right)
^{2}-m_{H^{\pm }}^{2}\right] }}\underset{a_{2}}{\underbrace{\left(
k^{2}-m_{l}^{2}\right) }}\underset{a_{3}}{\underbrace{\left[ \left(
k+p_{1}\right) ^{2}-m_{H^{\pm }}^{2}\right] }}}
\end{equation*}%
then%
\begin{eqnarray*}
\frac{1}{a_{1}^{1}a_{2}^{1}a_{3}^{1}} &=&\frac{\Gamma \left( 1+1+1\right) }{%
\Gamma \left( 1\right) \Gamma \left( 1\right) \Gamma \left( 1\right) }%
\int\limits_{0}^{1}dx\int\limits_{0}^{x}dy\frac{\left( x_{0}-x\right)
^{\left( 1-1\right) }\left( x-y\right) ^{\left( 1-1\right) }\left(
y-x_{3}\right) ^{\left( 1-1\right) }}{\left[ a_{3}\left( x_{0}-x\right)
+a_{2}\left( x-y\right) +a_{1}\left( y-x_{3}\right) \right] ^{3}} \\
&=&\Gamma \left( 3\right) \int\limits_{0}^{1}dx\int\limits_{0}^{x}dy\frac{1}{%
\left[ \left( \left( k+p_{1}\right) ^{2}-m_{H^{\pm }}^{2}\right) \left(
1-x\right) +\left( k^{2}-m_{l}^{2}\right) \left( x-y\right) +\left[ \left(
k+p_{2}\right) ^{2}-m_{H^{\pm }}^{2}\right] y\right] ^{3}}
\end{eqnarray*}%
where $x_{0}=1$ and $x_{3}=0$. Thus, the denominator can be written as 
\begin{eqnarray*}
&&\left[ \left( k+p_{1}\right) ^{2}-m_{H^{\pm }}^{2}\right] \left(
1-x\right) +\left( k^{2}-m_{l}^{2}\right) \left( x-y\right) +\left[ \left(
k+p_{2}\right) ^{2}-m_{H^{\pm }}^{2}\right] y \\
&=&k^{2}+2k\cdot \underset{b}{\underbrace{\left( p_{1}\left( 1-x\right)
+p_{2}y\right) }}+\underset{a^{2}}{\underbrace{\left( m_{H^{\pm
}}^{2}-m_{\nu }^{2}-m_{l}^{2}\right) x+\left( m_{l}^{2}+m_{\nu
}^{2}-m_{H^{\pm }}^{2}\right) y+m_{\nu }^{2}-m_{H^{\pm }}^{2}}} \\
&=&k^{2}+2k\cdot b+a^{2} \\
&=&k^{2}+P^{2}
\end{eqnarray*}%
where we use the transformation $k\rightarrow k-b$ in the last equation and 
\begin{equation*}
P^{2}=a^{2}-b^{2}=y^{2}m_{\nu }^{2}-2xym_{\nu }^{2}+\left( m_{\nu
}^{2}-m_{l}^{2}+m_{H^{\pm }}^{2}\right) y+x^{2}m_{\nu }^{2}+\left(
m_{l}^{2}-m_{\nu }^{2}-m_{H^{\pm }}^{2}\right) x+m_{H^{\pm }}^{2}
\end{equation*}%
in consequence, adding the corresponding terms to the integral with terms in
the numerator $1,k^{\mu }$ and $k^{\mu }k_{\nu }$ we obtain 
\begin{eqnarray*}
\Lambda _{2H^{\pm }1L}^{\alpha }\left( q,l\right) &=&-e\int \frac{d^{4}k}{%
\left( 2\pi \right) ^{4}}\left\{ \frac{\left( 2k^{\alpha }k_{\beta
}+p_{2}^{\alpha }k_{\beta }+p_{1}^{\alpha }k_{\beta }\right) A\gamma ^{\beta
}B+m_{l}\left( 2k^{\alpha }+p_{2}^{\alpha }+p_{1}^{\alpha }\right) AB}{\left[
\left( k+p_{1}\right) ^{2}-m_{H^{\pm }}^{2}\right] \left[ \left(
k+p_{2}\right) ^{2}-m_{H^{\pm }}^{2}\right] \left( k^{2}-m_{l}^{2}\right) }%
\right\} \\
&=&-e\frac{i}{16\pi ^{2}}\int\limits_{0}^{1}dx\int\limits_{0}^{x}dy\left[
\left( -\frac{2b^{\alpha }b^{\beta }}{P^{2}}+g^{\alpha \beta }\ln \frac{%
\Lambda ^{2}}{P^{2}}+\frac{p_{2\alpha }b^{\beta }}{P^{2}}+\frac{p_{1_{\alpha
}}b^{\beta }}{P^{2}}\right) A\gamma ^{\beta }B+\left( \frac{2b^{\alpha }}{%
P^{2}}-\frac{p_{2\alpha }}{P^{2}}-\frac{p_{1\alpha }}{P^{2}}\right) m_{l}AB%
\right]
\end{eqnarray*}%
expanding the terms $b^{\mu }$ and employing the Dirac equation%
\begin{eqnarray*}
\left( \gamma \cdot p_{1}-m\right) u\left( p_{1}\right) &=&0\Rightarrow \NEG%
{p}u\left( p_{1}\right) =mu\left( p_{1}\right) \\
\bar{u}\left( p_{2}\right) \left( \gamma \cdot p_{2}-m\right)
&=&0\Rightarrow \bar{u}\left( p_{2}\right) \NEG{p}_{2}=m\bar{u}\left(
p_{2}\right)
\end{eqnarray*}%
then%
\begin{eqnarray*}
&&\Lambda _{2H^{\pm }1L}^{\alpha }\left( q,l\right) \\
&=&-e\frac{i}{16\pi ^{2}}\int\limits_{0}^{1}dx\int\limits_{0}^{x}dy\left[
\left( -\frac{2b^{\alpha }b^{\beta }}{P^{2}}+g^{\alpha \beta }\ln \frac{%
\Lambda ^{2}}{P^{2}}+\frac{p_{2\alpha }b^{\beta }}{P^{2}}+\frac{p_{1_{\alpha
}}b^{\beta }}{P^{2}}\right) A\gamma ^{\beta }B+\left( \frac{2b^{\alpha }}{%
P^{2}}-\frac{p_{2\alpha }}{P^{2}}-\frac{p_{1\alpha }}{P^{2}}\right) m_{l}AB%
\right] \\
&=&-e\frac{i}{16\pi ^{2}}\int\limits_{0}^{1}dx\int\limits_{0}^{x}dy\left[ 
\frac{1}{P^{2}}\left( \left[ 
\begin{array}{c}
-p_{1\alpha }\left( 1-x\right) ^{2}-p_{2\alpha }y\left( 1-x\right)
-p_{1\alpha }\left( 1-x\right) y-p_{2\alpha }y^{2} \\ 
+\left( p_{2\alpha }\left( 1-x\right) +p_{2\alpha }y\right) \frac{1}{2}%
+\left( p_{1\alpha }\left( 1-x\right) +p_{1\alpha }y\right) \frac{1}{2}%
\end{array}%
\right] m_{\nu }\left( \left( ac+bd\right) +\left( bd-ac\right) \gamma
_{5}\right) \right. \right. \\
&&\left. \left. +m_{l}\left[ p_{1\alpha }\left( 1-x\right) +p_{2\alpha
}y-p_{2\alpha }\frac{1}{2}-p_{1\alpha }\frac{1}{2}\right] \left[ \left(
ac+bd\right) +\left( bd-ac\right) \gamma _{5}\right] \right) +\frac{1}{2}%
\gamma ^{\alpha }\left[ \left( bc+ad\right) -\left( bc-ad\right) \gamma _{5}%
\right] \ln \left( \frac{\Lambda ^{2}}{P^{2}}\right) \right]
\end{eqnarray*}%
and using the Gordon relation%
\begin{eqnarray*}
\overline{u}\left( p_{2\alpha }\right) \gamma _{\alpha }u\left( p_{1\alpha
}\right) &=&\frac{1}{2m_{\nu }}\overline{u}\left( p_{2\alpha }\right) \left[
\left( p_{2}+p_{1}\right) _{\alpha }+i\sigma ^{\alpha \mu }q_{\mu }\right]
u\left( p_{1\alpha }\right) \\
&\Rightarrow &\overline{u}\left( p_{2\alpha }\right) \left(
p_{2}+p_{1}\right) _{\alpha }u\left( p_{1\alpha }\right) =\overline{u}\left(
p_{2\alpha }\right) 2m_{\nu }\gamma _{\alpha }u\left( p_{1\alpha }\right) -%
\overline{u}\left( p_{2\alpha }\right) i\sigma ^{\alpha \mu }q_{\mu }u\left(
p_{1\alpha }\right)
\end{eqnarray*}%
finally the contribution for the EFF's with two charged Higgses and one
lepton can be represented by 
\begin{eqnarray*}
&&\Lambda _{2H^{\pm }1L}^{\alpha }\left( q,l\right) = \\
&&\frac{-ie}{16\pi ^{2}}\int\limits_{0}^{1}dx\int\limits_{0}^{x}dy\left[ 
\frac{1}{P^{2}}\left( \left( \left( m_{l}m_{\nu }-2m_{\nu }^{2}\right)
\gamma _{\alpha }+x\left( 6m_{\nu }^{2}-2m_{\nu }m_{l}\right) \gamma
_{\alpha }-4x^{2}m_{\nu }^{2}\gamma _{\alpha }+\left( m_{\nu }-\frac{1}{2}%
m_{l}\right) i\sigma ^{\alpha \mu }q_{\mu }\right. \right. \right. \\
&&\left. \left. +\left( m_{l}-3m_{\nu }\right) ix\sigma ^{\alpha \mu }q_{\mu
}+2x^{2}im_{\nu }\sigma ^{\alpha \mu }q_{\mu }\right) \left[ \left(
ac+bd\right) +\left( bd-ac\right) \gamma _{5}\right] +\frac{1}{2}\gamma
^{\alpha }\left[ \left( bc+ad\right) -\left( bc-ad\right) \gamma _{5}\right]
\ln \left( \frac{\Lambda ^{2}}{P^{2}}\right) \right]
\end{eqnarray*}

And for the case of two lepton and a charged Higgs $\left( 2L1H^{\pm
}\right) $ the general form of the contribution can be written as 
\begin{equation*}
\Lambda _{2L1H^{\pm }}^{\alpha }\left( q,l\right) =-e\int \frac{d^{4}k}{%
\left( 2\pi \right) ^{4}}\left\{ \frac{\left( aP_{L}+bP_{R}\right) \left( 
\NEG{k}+\NEG{p}_{1}+m_{l}\right) \gamma ^{\alpha }\left( \NEG{k}+\NEG%
{p}_{2}+m_{l}\right) \left( cP_{L}+dP_{R}\right) }{\left[ \left(
k+p_{1}\right) ^{2}-m_{l}^{2}\right] \left[ \left( k+p_{2}\right)
^{2}-m_{l}^{2}\right] \left( k^{2}-m_{H^{+}}^{2}\right) }\right\}
\end{equation*}%
expanding the numerator and using the same change as for the vertex $2H^{\pm
}1L$ we have 
\begin{eqnarray*}
&&\left. A\left( \NEG{k}+\NEG{p}_{1}+m_{l}\right) \gamma ^{\alpha }\left( 
\NEG{k}+\NEG{p}_{2}+m_{l}\right) B=\right. \\
&&\left( k_{\mu }k_{\beta }+p_{1\mu }k_{\beta }+k_{\mu }p_{2\beta }+p_{1\mu
}p_{2\beta }\right) A\gamma ^{\mu }\gamma ^{\alpha }\gamma ^{\beta
}B+m_{l}\left( k+p_{1}\right) _{\mu }A\gamma ^{\mu }\gamma ^{\alpha
}B+m_{l}\left( k+p_{2}\right) _{\beta }A\gamma ^{\alpha }\gamma ^{\beta
}B+m_{l}^{2}A\gamma ^{\alpha }B
\end{eqnarray*}%
and for the denominator we use the dimensional regularization method%
\begin{equation*}
\frac{1}{\underset{a_{1}}{\underbrace{\left[ \left( k+p_{2}\right)
^{2}-m_{l}^{2}\right] }}\underset{a_{2}}{\underbrace{\left( k^{2}-m_{H^{\pm
}}^{2}\right) }}\underset{a_{3}}{\underbrace{\left[ \left( k+p_{1}\right)
^{2}-m_{l}^{2}\right] }}}
\end{equation*}%
then%
\begin{equation*}
\frac{1}{a_{1}^{1}a_{2}^{1}a_{3}^{1}}=\Gamma \left( 3\right)
\int\limits_{0}^{1}dx\int\limits_{0}^{x}dy\frac{1}{\left[ \left( \left(
k+p_{1}\right) ^{2}-m_{l}^{2}\right) \left( 1-x\right) +\left(
k^{2}-m_{H^{\pm }}^{2}\right) \left( x-y\right) +\left( \left(
k+p_{2}\right) ^{2}-m_{l}^{2}\right) y\right] ^{3}}
\end{equation*}%
where $x_{0}=1$ and $x_{3}=0$. Hence, the denominator can be written as 
\begin{eqnarray*}
&&\left[ \left( k+p_{1}\right) ^{2}-m_{l}^{2}\right] \left( 1-x\right)
+\left( k^{2}-m_{H^{\pm }}^{2}\right) \left( x-y\right) +\left[ \left(
k+p_{2}\right) ^{2}-m_{l}^{2}\right] y \\
&=&k^{2}+2k\cdot \underset{b}{\underbrace{\left( p_{1}\left( 1-x\right)
+p_{2}y\right) }}+\underset{a^{2}}{\underbrace{\left( m_{l}^{2}-m_{\nu
}^{2}-m_{H^{\pm }}^{2}\right) x+\left( m_{H^{\pm }}^{2}+m_{\nu
}^{2}-m_{l}^{2}\right) y+m_{\nu }^{2}-m_{l}^{2}}} \\
&=&k^{2}+2k\cdot b+a^{2} \\
&=&k^{2}+P^{2}
\end{eqnarray*}%
where we use the transformation $k\rightarrow k-b$ in the last equation and 
\begin{equation*}
P^{2}=a^{2}-b^{2}=y^{2}m_{\nu }^{2}-\left( m_{H^{\pm }}^{2}-m_{\nu
}^{2}-m_{l}^{2}\right) y-2xym_{\nu }^{2}+x^{2}m_{\nu }^{2}-\left(
m_{l}^{2}+m_{\nu }^{2}-m_{H^{\pm }}^{2}\right) x+m_{l}^{2}
\end{equation*}%
therefore, adding the corresponding terms to the integral with terms $%
1,k^{\mu }$ and $k^{\mu }k_{\nu }$ in the numerator, we obtain 
\begin{eqnarray*}
&&\Lambda _{2L1H^{\pm }}^{\alpha }\left( q,l\right) \\
&=&-e\int \frac{d^{4}k}{\left( 2\pi \right) ^{4}}\left\{ \frac{\left( \left(
k_{\mu }k_{\beta }+p_{1\mu }k_{\beta }+k_{\mu }p_{2\beta }+p_{1\mu
}p_{2\beta }\right) A\gamma ^{\mu }\gamma ^{\alpha }\gamma ^{\beta
}B+m_{l}\left( k+p_{1}\right) _{\mu }A\gamma ^{\mu }\gamma ^{\alpha
}B+m_{l}\left( k+p_{2}\right) _{\beta }A\gamma ^{\alpha }\gamma ^{\beta
}B+m_{l}^{2}A\gamma ^{\alpha }B\right) }{\left[ \left( k+p_{1}\right)
^{2}-m_{l}^{2}\right] \left[ \left( k+p_{2}\right) ^{2}-m_{l}^{2}\right]
\left( k^{2}-m_{H^{\pm }}^{2}\right) }\right\} \\
&=&-e\frac{i}{16\pi ^{2}}\int\limits_{0}^{1}dx\int\limits_{0}^{x}dy\left[
\left( -\frac{b^{\mu }b^{\beta }}{P^{2}}+\frac{g^{\mu \beta }}{2}\ln \frac{%
\Lambda ^{2}}{P^{2}}+\frac{p_{1\mu }b^{\beta }}{P^{2}}+\frac{p_{2_{\beta
}}b^{\mu }}{P^{2}}-\frac{p_{1\mu }p_{2_{\beta }}}{P^{2}}\right) A\gamma
^{\mu }\gamma ^{\alpha }\gamma ^{\beta }B\right. \\
&&\left. +\left( \frac{b^{\mu }}{P^{2}}-\frac{p_{1\mu }}{P^{2}}\right)
m_{l}A\gamma ^{\mu }\gamma ^{\alpha }B+\left( \frac{b^{\beta }}{P^{2}}-\frac{%
p_{2\beta }}{P^{2}}\right) m_{l}A\gamma ^{\alpha }\gamma ^{\beta }B-\frac{1}{%
P^{2}}m_{l}^{2}A\gamma ^{\alpha }B\right]
\end{eqnarray*}%
expanding the terms $b^{\mu }$ and employing the Dirac equation we obtain%
\begin{eqnarray*}
&&\Lambda _{2L1H^{\pm }}^{\alpha }\left( q,l\right) \\
&=&-e\frac{i}{16\pi ^{2}}\int\limits_{0}^{1}dx\int\limits_{0}^{x}dy\left[
\left( -\frac{b^{\mu }b^{\beta }}{P^{2}}+\frac{g^{\mu \beta }}{2}\ln \frac{%
\Lambda ^{2}}{P^{2}}+\frac{p_{1\mu }b^{\beta }}{P^{2}}+\frac{p_{2_{\beta
}}b^{\mu }}{P^{2}}-\frac{p_{1\mu }p_{2_{\beta }}}{P^{2}}\right) A\gamma
^{\mu }\gamma ^{\alpha }\gamma ^{\beta }B+\left( \frac{b^{\mu }}{P^{2}}-%
\frac{p_{1\mu }}{P^{2}}\right) m_{l}A\gamma ^{\mu }\gamma ^{\alpha }B\right.
\\
&&\left. +\left( \frac{b^{\beta }}{P^{2}}-\frac{p_{2\beta }}{P^{2}}\right)
m_{l}A\gamma ^{\alpha }\gamma ^{\beta }B-\frac{1}{P^{2}}m_{l}^{2}A\gamma
^{\alpha }B\right] \\
&=&-e\frac{i}{16\pi ^{2}}\int\limits_{0}^{1}dx\int\limits_{0}^{x}dy\left[ 
\frac{1}{P^{2}}\left( \frac{1}{2}\left( x^{2}-2xy+2x+y^{2}-y\right) m_{\nu
}^{2}\gamma ^{\alpha }\left( \left( bc+ad\right) -\left( bc-ad\right) \gamma
_{5}\right) \right. \right. \\
&&+\left( \left( y-x+xy\right) \left( p_{2\alpha }+p_{1\alpha }\right)
-p_{2\alpha }y^{2}+p_{1\alpha }\left( x-y-x^{2}\right) \right) m_{\nu
}\left( \left( ac+bd\right) +\left( bd-ac\right) \gamma _{5}\right) \\
&&m_{l}m_{\nu }\gamma ^{\alpha }\left( \left( bc+ad\right) -\left(
bc-ad\right) \gamma _{5}\right) +m_{l}\left( p_{2\alpha }\left( y-1\right)
-xp_{1\alpha }\right) \left( \left( ac+bd\right) +\left( bd-ac\right) \gamma
_{5}\right) \\
&&\left. \left. -m_{l}^{2}\frac{1}{2}\gamma ^{\alpha }\left( \left(
bc+ad\right) -\left( bc-ad\right) \gamma _{5}\right) \right) -\frac{1}{2}%
\gamma ^{\alpha }\left( \left( bc+ad\right) -\left( bc-ad\right) \gamma
_{5}\right) \ln \frac{\Lambda ^{2}}{P^{2}}\right)
\end{eqnarray*}%
and using the Gordon relation like in the previous case, the contribution to
the EFF's with two leptons and one charged Higgs can be represented as 
\begin{eqnarray*}
&&\Lambda _{2L1H^{\pm }}^{\alpha }\left( q,l\right) = \\
&&\frac{-ie}{16\pi ^{2}}\int\limits_{0}^{1}dx\int\limits_{0}^{x}dy\left( 
\frac{1}{P^{2}}\left( 
\begin{array}{c}
\left( 2x^{2}m_{\nu }^{2}-\frac{1}{2}xm_{\nu }^{2}+m_{l}m_{\nu }-\frac{1}{2}%
m_{l}^{2}\right) \gamma ^{\alpha }\left( \left( bc+ad\right) -\left(
bc-ad\right) \gamma _{5}\right) \\ 
+\left( 2x\left( m_{\nu }^{2}-m_{l}m_{\nu }\right) -4m_{\nu
}^{2}x^{2}\right) \gamma _{\alpha }\left( \left( ac+bd\right) +\left(
bd-ac\right) \gamma _{5}\right)%
\end{array}%
\right. \right. \\
&&\left. \left. +\left( 2x^{2}m_{\nu }i\sigma ^{\alpha \mu }q_{\mu
}+m_{l}xi\sigma ^{\alpha \mu }q_{\mu }-xm_{\nu }i\sigma ^{\alpha \mu }q_{\mu
}\right) \left[ \left( ac+bd\right) +\left( bd-ac\right) \gamma _{5}\right]
\right) -\frac{1}{2}\gamma ^{\alpha }\left[ \left( bc+ad\right) -\left(
bc-ad\right) \gamma _{5}\right] \ln \frac{\Lambda ^{2}}{P^{2}}\right)
\end{eqnarray*}


\begin{thebibliography}{99}
\bibitem{Fukuda} Y. Fukuda et al. (Super-Kamiokande Collaboration) Phys.
Rev. Lett. 81, 1562 (1998).

\bibitem{Abe} K. Abe et al. (T2K Collaboration), PRL 112, 061802 (2014)

\bibitem{PDG} J. Beringer et al. (Particle Data Group), Phys. Rev. D86,
010001 (2013)

\bibitem{Mereghetti} S. Mereghetti, Astron. Astrophys. Rev. 15, 225 (2008);

\bibitem{Sher} G.C. Branco, P.M. Ferreira, L. Lavoura, M.N. Rebelo, Marc
Sher, Joao P. Silva, Theory and phenomenology of two-Higgs-doublet models,
Physics reports 516 (2012).

\bibitem{Pontecorvo} B. Pontecorvo, Journal of Experimental and Theoretical
Physics 26, 984 (1983);

\bibitem{Maki} Z. Maki, M. Nakagawa, S. Sakata, Progress of Theoretical
Physics 28, 870, (1962);

\bibitem{Nova} M. Nowakowski, E. A. Paschos, J. M. Rodriguez, Eur. J. Phys.
26 (2005) 545

\bibitem{broggini} Broggini, C., C. Giunti, and A. Studenikin,
Electromagnetic properties of neutrinos, Adv. High Energy Phys. 2012,
459526, arXiv:1207.3980 [hep-ph]

\bibitem{Giunti} Carlo Giunti, Alexander Studenikin. Neutrino
electromagnetic properties,[arXiv:1006.1502v1 [hep-ph]].

\bibitem{lepton} Lepton dipole moments (ed B. Roberts, J. Marciano) world
Scientific, (2010);

\bibitem{2HDM} J. Gunion, H. Haber, G. Kane, S. Dawson. Higgs hunters guide,
Perseus publishing. (2000);

\bibitem{Diaz} R. A. Diaz, Ph.D. Thesis [arXiv: hep-ph/0212237].

\bibitem{Carena} M. Carena, H. Haber. Higgs Boson Theory and Phenomenology,
Prog. Part. Nucl. Phys. 50, 63 (2003). [arXiv:hep/ph/0208209]

\bibitem{Fujikawa} K. Fujikawa, R. Shrock, The Magnetic Moment of a Massive
Neutrino and Neutrino Spin Rotation, Phys. Rev. Lett. 45 (1980) 963.

\bibitem{Akeroyda} A.G. Akeroyd and F. Mahmoudi, Constraints on charged
Higgs bosons from $D_{s}^{\pm }\rightarrow \mu ^{\pm }\nu _{\mu }$ and $%
D_{s}^{\pm }\rightarrow \tau ^{\pm }\nu _{\tau }$, JHEP04(2009)121

\bibitem{Mahmoudi} F. Mahmoudi and O. Stål, Flavor constraints on
two-Higgs-doublet models with general diagonal Yukawa couplings, Phys. Rev.
D 81, 035016, (2010)

\bibitem{Wong} H. Wong, et al., A Search of Neutrino Magnetic Moments with a
High-Purity Germanium Detector at the Kuo-Sheng Nuclear Power Station,
Phys.Rev. D75 (2007) 012001.

\bibitem{Gemma} A. Beda, V. Brudanin, V. Egorov, D. Medvedev, V. Pogosov, et
al., Gemma experiment: The results of neutrino magnetic moment search,
Phys.Part.Nucl.Lett. 10 (2013) 139-143.

\bibitem{Auerbach} L. B. Auerbach, R. L. Burman, D. O. Caldwell et al.,
Measurement of electron-neutrino electron elastic scattering, Physcial
Review D, vol. 63, no. 11, 11 pages, 2001.

\bibitem{Montanino} D. Montanino, M. Picariello, and J. Pulido, Probing
neutrino magnetic moment and unparticle interactions with Borexino, Physical
Review D, vol. 77, no. 9, Article ID 093011, 9 pages, 2008.

\bibitem{Schwienhorst} R. Schwienhorst, D. Ciampa, C. Erickson et al., A new
upper limit for the tau-neutrino magnetic moment, Physics Letters B, vol.
513, no. 1-2, pp. 23--29, 2001.
\end{thebibliography}
\end{document}